\documentclass[a4paper,11pt]{article}
\pdfoutput=1
\usepackage{jheppub}

\usepackage[T1]{fontenc}

\newcommand{\nocontentsline}[3]{}
\newcommand{\tocless}[2]{\bgroup\let\addcontentsline=\nocontentsline#1{#2}\egroup}

\usepackage[utf8]{inputenc}	
\usepackage{framed}

\usepackage{caption}
\usepackage{subcaption}
\usepackage{booktabs}

\usepackage{amsmath}
\usepackage{amssymb}
\usepackage{amscd}
\usepackage{mathrsfs}
\usepackage{graphicx}
\usepackage{slashed}
\usepackage{mathbbol}
\usepackage{multirow}
\usepackage{array}
\usepackage{afterpage}

\numberwithin{equation}{section}

\usepackage{fmdg}

\newcommand{\tr}{\ensuremath{\mathrm{tr}}}

\newcommand{\Eqref}[1]{Eq.~(\ref{#1})}

\newcommand{\Figref}[1]{Fig.~\ref{#1}}
\newcommand{\Tabref}[1]{Tab.~\ref{#1}}
\newcommand{\Secref}[1]{Sec.~\ref{#1}}
\newcommand{\Appref}[1]{Appendix \ref{#1}}

\newcommand{\eV}{\ensuremath{\,\mathrm{eV}}}

\newcommand{\hc}{\ensuremath{\mathrm{h.c.}}}

\definecolor{MS}{rgb}{0,0,1}

\usepackage{cancel}

\title{\boldmath Testing Radiative Neutrino Mass Models \\at the LHC}

\author[a]{Yi Cai,}
\author[a]{Jackson D.~Clarke,}
\author[b]{Michael A.~Schmidt,}
\author[a]{and Raymond R.~Volkas}

\affiliation[a]{ARC Centre of Excellence for Particle Physics at the Terascale,\\
School of Physics, The University of Melbourne, Victoria 3010, Australia}
\affiliation[b]{ARC Centre of Excellence for Particle Physics at the Terascale,\\
School of Physics, The University of Sydney, NSW 2006, Australia}

\emailAdd{yi.cai@unimelb.edu.au}
\emailAdd{j.clarke5@pgrad.unimelb.edu.au}
\emailAdd{m.schmidt@physics.usyd.edu.au}
\emailAdd{raymondv@unimelb.edu.au}

\abstract{
The Large Hadron Collider provides us new opportunities to search for the origin of neutrino mass.
Beyond the minimal see-saw models a plethora of models exist which realise
neutrino mass at tree- or loop-level, and it is important to be sure
that these possibilities are satisfactorily covered by searches.
The purpose of this paper is to advance a systematic approach to this problem.
Majorana neutrino mass models can be organised by SM-gauge-invariant
operators which violate lepton number by two units.
In this paper we write down the minimal ultraviolet completions for all of the mass-dimension 7 operators.
We predict vector-like quarks, vector-like leptons, scalar leptoquarks, a charged scalar, a scalar doublet, and a scalar quadruplet, 
whose properties are constrained by neutrino oscillation data.
A detailed collider study is presented for $\mathcal{O}_3=LLQ\bar{d}H$ and
$\mathcal{O}_8=L\bar{d}\bar{e}^\dagger\bar{u}^\dagger H$ completions with a 
vector-like quark $\chi\sim (3,2,-\frac{5}{6})$ and a leptoquark $\phi\sim (\bar{3},1,\frac{1}{3})$.
The existing LHC limits extracted from searches for
vector-like fermions and sbottoms/stops  are  $m_\chi\gtrsim 620 \ \rm{GeV}$ and $m_\phi \gtrsim 600 \ \rm{GeV}$.
}

\begin{document}

\maketitle
\flushbottom

\setcounter{footnote}{0}

\section{Introduction}
Neutrino oscillation experiments~\cite{Fukuda:1998mi,Apollonio:1999ae,Ahmad:2002jz,Eguchi:2002dm,Ahmed:2003kj,Ahn:2006zza,Michael:2006rx,Abe:2011sj,Adamson:2011qu,Abe:2011fz,An:2012eh,Ahn:2012nd,Abe:2012tg,An:2013uza} have established that neutrinos change flavour in a manner that is perfectly consistent with the standard mechanism: the flavour eigenstates are unitary superpositions of non-degenerate mass eigenstates that, after creation, evolve in time as free particles.  The origin of the required neutrino masses and mixings continues to be one of the outstanding problems in particle physics.  The neutrinos have unusually small masses (sub-eV) and the leptonic unitary (PMNS) mixing matrix~\cite{Maki:1962mu} is of a different qualitative form from the quark analogue.  These observations, especially the former, strongly suggest that the neutrino mass generation mechanism is different from that of the charged fermions.  A key distinguishing feature is that neutrinos may be Majorana fermions, the case we consider in this paper.

A much-studied possibility is that neutrinos may pick up mass at tree-level through one of the see-saw mechanisms~\cite{Minkowski:1977sc, Yanagida:1980, Glashow:1979vf, Gell-Mann:1980vs, Mohapatra:1980ia, Magg:1980ut, Schechter:1980gr,Wetterich:1981bx, Lazarides:1980nt, Mohapatra:1980yp, Cheng:1980qt, Foot:1988aq}. 
  Another generic possibility, the focus of this work, is that the origin is radiative, at $1$- to $3$-loop order~\cite{Zee:1980ai, Cheng:1980qt, Zee:1985id, Babu:1988ki, Krauss:2002px}. 
One reason to be interested in such models is that the new physics required may be searched for, or non-trivially constrained, at the Large Hadron Collider (LHC)\footnote{One possible argument for a low scale of neutrino mass generation is classical scale invariance~\cite{Coleman:1973jx}, which has been studied in Refs.~\cite{Foot:2007ay,Lindner:2014oea}. In general the seesaw framework is difficult to test at the LHC, but there are some regions of parameter space which might be testable at colliders. See e.g.~Ref.~\cite{Kersten:2007vk,Pilaftsis:1991ug,Dev:2012sg}.}, in addition to having flavour-violation signatures. 

A challenge is that there are many viable radiative models, and one wishes to study them in as generic and inclusive a way as possible.  One very good way to approach this task is to begin with gauge-invariant effective operators that violate lepton-number by two units ($\Delta L =2$), constructed out of standard model (SM) fields~\cite{Babu:2001ex,deGouvea:2007xp,Angel:2012ug}.  These operators, which Babu and Leung~\cite{Babu:2001ex} systematically classified for mass dimensions 5, 7, 9 and 11, produce vertices that feature in loop-level graphs generating Majorana masses (and mixing angles and phases).  By opening up the operators in all possible ways subject to some minimality assumptions, one may in principle construct all candidate renormalizable models that yield radiative Majorana neutrino masses consistent with those assumptions~\cite{Angel:2012ug}. Following Ref.~\cite{Angel:2012ug} we restrict ourselves to tree-level UV completions.
Alternatively neutrino mass models can be classified according to a subset of $\Delta L=2$ operators of the form $LLHH (H^\dagger H)^n$, which has been pursued in Refs.~\cite{Bonnet:2009ej,Bonnet:2012kz,Krauss:2013gy}.

The purpose of this paper is twofold.  First, we write down the candidate models implied by opening up all of the dimension-7 (D7) operators in the Babu-Leung list, subject to the following minimality assumptions: (a) The gauge symmetry is that of the SM only, and effective operators containing gauge fields are excluded from consideration. (b) The exotic particles that are integrated-out to produce the effective operators are either scalars, vector-like fermions or Majorana fermions.  The Appendix is a compendium of all of the candidate models.  

The second purpose is to do a detailed study of the LHC constraints and signatures, taking account of flavour-violation constraints in the process.  This study raises its own challenges, because each model has its special features.  We approach this by first listing the quantum numbers of all the exotic scalars and fermions that appear in at least one of the D7 models, and then summarising the existing constraints from ATLAS and CMS.  In a second stage, we analyse one of the models in detail to determine the precise LHC reach and constraints.

The remainder of this paper is structured as follows.
In Sec.~\ref{sec:uvd7} we study the minimal UV completions of the D7 operators 
and list all the exotic particles grouped according to their completion topologies.
The details of these UV completions are given in App.~\ref{app:uvcompletion}.
Searches for these exotic particles will be generally discussed in 
Sec.~\ref{sec:lhcsearches} including the production mechanisms, decay patterns and the searching strategies. 
Experimental limits will be presented if there are dedicated searches. We then present a detailed analysis
of a specific model in Sec.~\ref{sec:example}. Constraints
from neutrino mass generation and flavour physics are explored,
and limits from LHC searches are derived.
Finally, Sec.~\ref{sec:con} is devoted to the conclusions.

\mathversion{bold}
\section{Minimal UV Completion of D7 $\Delta L=2$ Operators}
\label{sec:uvd7}
\mathversion{normal}

In Weyl-spinor notation, the D7 operators of interest, using the numbering system of Babu-Leung~\cite{Babu:2001ex}, are
\begin{align}
 \mathcal{O}_2 &= LLL\bar{e}H ,&
 \mathcal{O}_3 &= LLQ\bar{d}H , &
 \mathcal{O}_4 &= LLQ^\dagger \bar{u}^\dagger H ,&
 \mathcal{O}_8 &= L\bar{d}\bar{e}^\dagger \bar{u}^\dagger H,&
 \label{eq:d7operators}
\end{align}
and the Weinberg-like operator
\begin{align}
\mathcal{O}_1^\prime &= LL \tilde HHHH .
\end{align}
The pertinent part of the SM Lagrangian is
\begin{align}
\mathcal{L}_{SM,Y}&= Y_e L \bar{e} \tilde{H} + Y_u Q \bar{u} H + Y_d Q \bar{d} \tilde{H} +\hc\;,
\end{align}
where $\tilde{H}=i\tau_2 H^*$ is the charge conjugate of $H$.
The Weinberg-like operator $\mathcal{O}_1^\prime$ has not been explicitly shown in the list of Babu-Leung~\cite{Babu:2001ex}, but has been studied in Refs.~\cite{Bonnet:2009ej,Krauss:2013gy}. Note that this operator always induces the usual Weinberg operator $\mathcal{O}_1=LLHH$ by connecting the two external legs $H$ and $\tilde H$ via a Higgs boson to form a Higgs loop. This contribution dominates if the scale of new physics is large, much above the TeV-scale.

We will study minimal ultraviolet (UV) completions of these D7 operators using scalars and fermions, 
following the programme set out in Ref.~\cite{Angel:2012ug}. Hence we do not include models with new gauge bosons.
Also, we only consider models which do not generate the dimension-5 Weinberg operator at tree-level. Hence we remove models in which one of the three seesaw mechanisms may operate, i.e. models containing SM singlet fermions, electroweak (EW) triplet scalars with unit hypercharge, and EW triplet fermions.

\begin{figure}[tp]
\begin{minipage}{0.5\linewidth}\centering
\FMDG{FFFF}
\captionof{figure}{Scalar-only extension.}
\label{fig:OnlyScalarExtension}
\end{minipage}
\begin{minipage}{0.5\linewidth}\centering
\begin{tabular}{ccl}
\toprule
Scalar & Scalar & Operator\\\midrule
  $(1, 2, \frac12)$ &   $(1, 1, 1)$ & $\mathcal{O}_{2,3,4}$ \cite{Zee:1980ai} \\
  $(3, 2, \frac16)$ &   $(3,1,-\frac13)$ & $\mathcal{O}_{3,8}$ \cite{Babu:2001ex,Babu:2010vp}\\
    $(3, 2, \frac16)$ &   $(3,3,-\frac13)$ & $\mathcal{O}_3$\\
\bottomrule
\end{tabular}
\captionof{table}{Topology of \Figref{fig:OnlyScalarExtension}.}
\label{tab:completionstab1}
\end{minipage}

\begin{minipage}{0.5\linewidth}\centering
\FMDG{FHFF}
\captionof{figure}{Extension by a scalar and a fermion.}
\label{fig:FermionScalarExtension}
\end{minipage}
\begin{minipage}{0.5\linewidth}\centering
\begin{tabular}{ccl}
\toprule
Dirac fermion & Scalar& Operator\\\midrule
  $(1, 2, -\frac32)$ &   $(1, 1, 1)$  &  $\mathcal{O}_2$ \\
  $(3, 2, -\frac56)$ &   $(1, 1, 1)$ & $\mathcal{O}_3$\\
  $(3, 1, \frac23)$ &   $(1, 1, 1)$ & $\mathcal{O}_3$\\
  $(3, 1, \frac23)$ &   $(3,2,\frac16)$ & $\mathcal{O}_3$ \cite{Babu:2011vb}\\
  $(3, 2, -\frac56)$ &   $(3,1,-\frac13)$ & $\mathcal{O}_{3,8}$$^\ast$ \\
  $(3, 2, -\frac56)$ &   $(3,3,-\frac13)$ & $\mathcal{O}_3$\\
  $(3, 3, \frac23)$ &   $(3,2,\frac16)$ & $\mathcal{O}_3$\\
  $(3, 2, \frac76)$ &   $(1, 1, 1)$ &$\mathcal{O}_4$\\
  $(3, 1, -\frac13)$ &   $(1, 1, 1)$ & $\mathcal{O}_4$\\
  $(3, 2, \frac76)$ &   $(3,2,\frac16)$ &  $\mathcal{O}_8$\\
  $(1,2,-\frac12)$ &   $(3,2,\frac16)$ &$\mathcal{O}_8$\\
\bottomrule
\end{tabular}
\captionof{table}{Topology of \Figref{fig:FermionScalarExtension}.
\label{tab:completionstab2}
The completion marked with a $^\ast$ is studied in detail in Sec.~\ref{sec:example}.}
\end{minipage}

\begin{minipage}{0.5\linewidth}\centering
\FMDG{FHHHH}
\captionof{figure}{Extension by a scalar and a fermion.}
\label{fig:FHHHH}
\end{minipage}
\begin{minipage}{0.5\linewidth}\centering
\begin{tabular}{ccl}
\toprule
Dirac fermion & Scalar& Operator\\\midrule
  $(1, 3, -1)$ &   $(1, 4, \frac32)$  & $\mathcal{O}_1^\prime$\cite{Babu:2009aq}\\
\bottomrule
\end{tabular}
\captionof{table}{Topology of \Figref{fig:FHHHH}.}
\label{tab:completionstab3}
\end{minipage}
\end{figure}

We group the completions by topology in Figs.~\ref{fig:OnlyScalarExtension}--\ref{fig:FHHHH} and Tables~\ref{tab:completionstab1}--\ref{tab:completionstab3}, where
quantum numbers are given with respect to SU(3)$_c\times$SU(2)$_L\times$U(1)$_Y$.
Details are left to App.~\ref{app:uvcompletion}.
The contents of Tables~\ref{tab:completionstab1}--\ref{tab:completionstab3} constitute a workable list of exotic particles relevant to D7 radiative neutrino mass models which may be searched for at the LHC. 

It turns out that the operators $\mathcal{O}_2$ and $\mathcal{O}_{3b}$ lead to one-loop models, while the others only admit two-loop models.
Generally for models with scalar leptoquarks and vector-like fermions, the radiatively generated neutrino mass is proportional to the quark or
lepton mass in the loop (we will show this in detail in Sec.~\ref{sec:example}). Thus we will only consider the mixing
between the third generation quarks or leptons and the exotic fermions, as the third generation masses dominate the neutrino mass matrix unless there is an unnatural flavour structure for the various coupling constants.

\afterpage{\clearpage}

\section{LHC Searches}
\label{sec:lhcsearches}

The completions listed in Tables~\ref{tab:completionstab1}--\ref{tab:completionstab3} each contain two fields beyond the SM, including vector-like quarks, vector-like leptons, scalar leptoquarks, charged scalars, EW scalar doublets and EW scalar quadruplets. 
In this section we discuss the pertinent LHC searches and limits for the lightest of these exotic fields.\footnote{Note, however, that the completions generically predict more complex cascade decays if allowed by the relevant couplings and phase space.}
We present a brief discussion of their production mechanisms and possible decay channels. 
A dedicated search at the LHC may or may not already exist. 
For those that exist, we list or reference the most stringent limits; 
these limits are generally functions of decay branching ratios which are treated as free parameters.
For those that do not exist, we list what would be the relevant LHC search according to the appropriate final state(s). 

We would like to emphasise that one of the advantages to our approach is its predictivity. 
The exotic particles are required to not only conform to existing flavour constraints, but also to fit low energy neutrino measurements.
As a result it is common in these neutrino mass generation models to be able to predict the decay patterns of the exotic particles. 
Then for a specific model it is possible to extract the limit based on the decay patterns,
either from existing searches, as we shall see in Sec.~\ref{sec:example:vlq}, or by carefully recasting relevant LHC searches, 
as in Sec.~\ref{sec:detail:leptoquark}.

\subsection{Vector-Like Quarks}
\label{subsec:vectorlikequarks}

In the minimal UV completion of the D7 operators we introduced five vector-like quarks,
\begin{center}
\begin{tabular}{ccccc}
\toprule
$B$ & $T$ & $(BY)$ & $(XT)$ & $(XTB)$\\ 
\midrule
$(3,1,-\frac{1}{3})$ & $(3,1,\frac{2}{3})$ & $(3,2,-\frac{5}{6})$ & $(3,2,\frac{7}{6})$ & $(3,3,\frac{2}{3})$\\
\bottomrule
\end{tabular}
\end{center}
where the names of the fermions follow the conventions in the literature~\cite{AguilarSaavedra:2009es}. 

Vector-like quarks are well-studied by the LHC collaborations.
They can be pair-produced in $pp$ collisions via gluon fusion and quark-antiquark annihilations.
They can also be singly produced in association with two extra quarks via $t$-channel processes involving a $W$ or $Z$ boson.
Single production depends on the mixing between the heavy fermions and the third generation quarks
and also the generalised CKM matrix, and can be dominant for large vector-like quark masses and large mixings~\cite{AguilarSaavedra:2009es}.
So far, collider studies have focused on the more model-independent pair-production channel.

The decay channels for the singlet $B$ and $T$ are
\begin{align}
B: \qquad B&\to W^- t,& B&\to Z b,&  B&\to H b  \ ,\\
T: \qquad T&\to W^+ b,& T&\to Z t,&  T&\to H t   \ ,
\end{align}
the branching fractions of which are determined by the masses of the heavy fermions and also the mixings between 
the heavy fermions and the third generation quarks together with the generalised CKM matrix.
The decays for the doublets and the triplet are determined by the mass spectrum and their weak coupling to the $W$ 
and $Z$ bosons. In general, the mass splitting among the components fields is suppressed by the mixing angles between
the SM quarks and the heavy quarks, which in turn suppresses the decays between the component fields.
For the two doublets $(BY)$ and $(XT)$, the possible decay channels are 
\begin{align}
(BY):\qquad Y&\to W^- b,&  B&\to Z b, & B&\to H b \ ,\\
(XT): \qquad X&\to W^+ t,&  T&\to Z t, & T&\to H t \ .
\end{align}        
For the triplet $(XTB)$, the possible decay channels are
\begin{eqnarray}
 (XTB): && X \to W^+ t \ ,\\
        && T\to W^+ b, \qquad T\to Z t, \qquad T\to H t \ ,\\
        && B\to W^- t ,\qquad B \to Z b, \qquad  B\to H b \ . 
\end{eqnarray} 
Note that the heavy $T$ and $B$ in the triplet $(XTB)$ also decay to $W$ like the singlet 
$T$ and $B$. Assuming only strong pair production and the same decay branching ratios, 
the limits we can set on the masses are the same for the singlet and the triplet $T$ and $B$.      

Both ATLAS~\cite{ATLAS:2013ima,TheATLAScollaboration:2013jha,TheATLAScollaboration:2013oha,TheATLAScollaboration:2013sha}
and CMS~\cite{CMS:2013una,CMS:2012hfa,CMS:2013zea,Chatrchyan:2013uxa,Chatrchyan:2013wfa} 
have performed searches for vector-like quarks, although there is no dedicated search for $Y$ so far.     
We list the limits from the CMS searches in Table~\ref{tab:lhclimits},
to be used later in extracting limits for the vector-like quarks we are interested in. 
\begin{table}[tp]
\centering
\begin{tabular}{cccc}
\toprule
Particle &  $T$ &  $B$ &  $X$\\
\midrule
Lower Mass Limit (GeV) & $687-782$~\cite{Chatrchyan:2013uxa} & $520-785$~\cite{CMS:2013una,CMS:2012hfa,CMS:2013zea} & $800$~\cite{Chatrchyan:2013wfa} \\
\bottomrule
\end{tabular}
\caption{The lower limits on the masses of the vector-like quarks from CMS.}
\label{tab:lhclimits}
\end{table} 

In practice, extracting the relevant limits from these dedicated searches involves calculation of 
the decay branching ratios of the exotic particle with the constraints from neutrino masses and mixings.
With the specific decay branching ratios, we will be able to pin down the limits by interpolation 
as shown in Sec.~\ref{sec:example:vlq} for $B$.

\subsection{Vector-Like Leptons}
\label{subsec:vectorlikeleptons}

Our completions also introduced three vector-like fermions which are not charged under $SU(3)_c$,
\begin{center}
\begin{tabular}{cccc}
\toprule
$E$ & $(NE)$ & $(ED)$ & $(NED)$ \\ 
\midrule
$(1,1,-1)$  & $(1,2,-\frac{1}{2})$ & $(1,2,-\frac{3}{2})$  & $(1,3,-1)$ \\
\bottomrule
\end{tabular}
\end{center}
Vector-like leptons which are singlets or doublets of SU(2)$_L$ have been thoroughly studied in the recent literature~\cite{Altmannshofer:2013zba, Falkowski:2013jya, Dermisek:2014qca}, while the triplet has been mentioned in the context of minimal dark matter~\cite{Cirelli:2005uq}.

The dominant production mechanism for these exotic leptons is Drell-Yan pair production. A pair of 
different-charge vector-like leptons can be subdominantly produced through an $s$-channel $W$. 
The vector-like leptons can also be singly produced with a SM lepton via $s$-channel $W$, $Z$ or Higgs.    
The subsequent decays of the vector-like leptons depends on the mass spectrum and mass mixing parameters.
Similarly to the vector-like quarks, the mass splittings among the component fields of the heavy fermions is 
suppressed and the possible decay channels are
\begin{eqnarray}
 E: &&\qquad E\to W^- \nu_\tau,  \qquad E\to Z\tau, \qquad E\to H\tau \ ,\\
 (NE):&&\qquad E\to W^- \nu_\tau, \qquad  N\to Z\nu_\tau, \qquad N\to H\nu_\tau \ , \\
 (ED):&& \qquad D\to W^- \tau, \qquad E\to Z\tau, \qquad E\to H\tau \ ,\\
 (NED):&& \qquad D\to W^- \tau,\\ 
 &&\qquad E\to W^- \nu ,\qquad E\to Z\tau , \qquad E\to H\tau ,\\
&&\qquad N \to W^+ \tau, \qquad N\to Z\nu, \qquad N\to H\nu \ .
\end{eqnarray}   
Thus pair-produced $N$, $E$ or $D$ will produce final states with a pair of bosons, $\tau$ lepton(s) and/or large missing transverse energy.
In general these models are constrained by the LHC searches for final states with $\tau$ lepton(s) and/or missing transverse energy together with 
leptons and/or jets.
So far there are no dedicated searches for these vector-like leptons at ATLAS and CMS. 
However, searches for multi-lepton plus missing transverse energy final states, including  some supersymmetry (SUSY) searches 
for sleptons or charginos~\cite{Aad:2014vma, Aad:2014yka, Aad:2014nua, Aad:2014iza, Khachatryan:2014qwa}, 
could be used to derive the bounds on vector-like leptons.   
For example, Ref.~\cite{Altmannshofer:2013zba} has studied the pair-production of $D$ fermions
which decay to light leptons or a combination of light leptons and at least one $\tau$, 
which  constrains the mass of vector-like leptons to be heavier than $460\; \rm{GeV}$ and $320\; \rm{GeV}$
respectively. 

\subsection{Leptoquarks}
\label{subsec:leptoquarks}

There are only five scalar leptoquark candidates whose interactions with SM fermions can be described 
by a dimensionless, SM gauge-invariant, baryon- and lepton-number conserving Lagrangian \cite{Buchmuller:1986zs}. 
Three of these leptoquarks have been introduced in our UV completions: 
\begin{center}
\begin{tabular}{ccc}
\toprule
$S_1$ & $\tilde{R}_2$ & $S_3$  \\
\midrule
$(\bar{3},1,\frac{1}{3})$  & $(3,2,\frac{1}{6})$ & $(\bar{3},3,\frac{1}{3})$ \\
\bottomrule
\end{tabular}
\end{center}
A recent systematic study of models of neutrino mass generation with leptoquarks can be found in Ref.~\cite{AristizabalSierra:2007nf}. 
Searches at the LHC assume simplified models in which the leptoquarks 
couple exclusively to leptons and quarks 
of a single generation in a chiral interaction. This assumption is made in order to not induce unacceptable flavour-changing currents 
or lepton-flavour violating four-fermion interactions. The most stringent of these limits come from meson mixing in the quark sector leading to a limit on the scale of the four-fermion interaction up to $10^8$ GeV (see e.g. Ref.~\cite{Cirigliano:2013lpa}). The limits in the lepton sector are generally weaker. The most stringent limits are from the $\mu\to e$ transition with Br$(\mu\to e \gamma)< 5.7\times10^{-13}$ \cite{Adam:2013mnn}, Br$(\mu\to eee)<10^{-12}$ \cite{PDG:2012}, and Br$(\mu \mathrm{Au}\to e\mathrm{Au})<7\times 10^{-13}$ \cite{PDG:2012}. As the limits in the lepton sector are weaker, it is possible to relax the strong assumption of an exclusive coupling to one generation in the lepton sector.
The collaborations make the further underlying assumption that the couplings are small enough so that one may only consider pair production governed by the leptoquark colour charge.

After pair production, final states of interest for first (second) generation leptoquarks are $ejej,ej\nu j,\nu j \nu j$ ($e\leftrightarrow \mu$).
Limits are set on $(m_{LQ},\beta)$ parameter space, where $\beta$ is the branching ratio to the charged lepton and quark
\cite{ATLAS:2012aq,Aad:2011ch,CMS-PAS-EXO-12-041,CMS-PAS-EXO-12-042}.
In practice the $\nu j \nu j$ state is not considered, although it would be constrained by SUSY searches for $\ge 2j+\slashed{E}_T$.
Searches for third generation leptoquarks consider only single decay hypotheses: $\tau b$, $\tau t$, $\nu b$, $\nu t$
\cite{ATLAS:2013oea,Khachatryan:2014ura,CMS-PAS-EXO-13-010,Chatrchyan:2012st}. 
The latter two are also covered by pair-produced sbottom and stop searches in the $m_{LSP}\rightarrow 0$ limit (LSP means the lightest supersymmetric particle). We discuss the $\nu b$ and $\ell t$ channels in more detail in Sec.~\ref{sec:example}.

\subsection{Charged Scalar}
\label{subsec:chargedscalar}

The charged scalar introduced in our completions is $\phi\sim (1,1,1)$. 
It couples to lepton bilinears and decays as
\begin{equation}
\phi \rightarrow \nu_i l_j^+,
\end{equation}  
which, after pair-production, results in the signature of two opposite-sign leptons and $\slashed{E}_T$ at the LHC.
There is also no dedicated search for such a scalar so far. 
However the same signature has been used to search for direct slepton-pair and chargino-pair production at the LHC~\cite{CMS:2013bda, Aad:2014yka, Aad:2014vma}. 
The limits of the SUSY search are given in a slepton- or chargino-neutralino mass plane, 
from which the limit of the charged scalar can be extracted by 
recasting the searches with the limit of $m_{LSP}\to 0$ and taking into account the different branching ratios. 

\subsection{EW Scalar Doublet}

\label{subsec:ewscalardoublet}
The only EW scalar doublet introduced is another Higgs doublet, $H\sim (1,2,1/2)$.
It can be decomposed as
\begin{equation}
H=\left(H^+, \frac{H^0 + i A^0}{\sqrt{2}}\right)^T.
\end{equation}
There has been extensive study of the SM extension with a second Higgs doublet (2HDM),
and analyses of the general 2HDM after LHC~Run~1 have been presented in recent studies~\cite{Altmannshofer:2012ar, Bai:2012ex, Celis:2013ixa, Craig:2013hca}.
The EW scalar doublet in the UV completions of $\mathcal{O}_{2,3,4}$ in general should have the same
couplings to the SM particles as in a 2HDM without imposed new symmetries. When we study the neutrino mass generation of a specific model, however,
it is possible to switch off many of the couplings without spoiling the generation of appropriate neutrino masses and mixings.
Thus the decay of the EW scalar doublet is fairly model-dependent and interpretation of LHC searches should be handled with caution.  

The combined results of the search for the SM Higgs at the LHC have been reported 
in Refs.~\cite{Chatrchyan:2012tx, Aad:2012tfa, Chatrchyan:2012ufa}. So far in the mass range of 127--600 GeV the SM Higgs
has been excluded at 95\% CL. Based on these limits, one can in principle draw a limit on the mass of $H^0$ by 
recasting the neutral Higgs search with rescaled decay branching ratios.
LEP has set a limit of $79.3\ \rm{GeV}$ on the charged Higgs mass 
assuming $\mathrm{Br}(H^+\rightarrow \tau^+\nu)+ \mathrm{Br}(H^+\rightarrow c\bar{s}) = 1$ in the framework of a 2HDM \cite{Heister:2002ev}.
Charged Higgs searches at the LHC are categorised by the mass of the charged Higgs.
The light charged Higgs, $m_{H^+}<m_t$, is mainly searched for through
$t\bar{t}$ pair production with the subsequent decay $t\rightarrow H^+ b$~\cite{Aad:2012rjx, Aad:2012tj,
Chatrchyan:2012vca, Aaltonen:2011aj, Aaltonen:2009ke, Abazov:2009wy, Abazov:2009aa, Abazov:2009ae}.
The heavy charged Higgs, on the other hand, is mainly searched for in the singly produced channel with
the subsequent decay $H^+\rightarrow t\bar{b}$~\cite{Abazov:2008rn}. These searches are under some
specific theoretical frameworks and can be reinterpreted with careful conversion of the parameters.

\subsection{EW Scalar Quadruplet}
The EW scalar quadruplet can be decomposed as
\begin{equation}
\phi = \left( \phi^{+++}, \phi^{++}, \phi^+, \frac{\phi^0+i A^0}{\sqrt{2}}\right)^T
\end{equation}
and contains a neutral scalar $\phi^0$ which mixes with the Higgs, a pseudo-scalar $A^0$ and three complex scalars. This scalar quadruplet has also been mentioned in the context of minimal dark matter~\cite{Cirelli:2005uq}, whose mass spectrum can be non-degenerate depending on the values of the parameters in the scalar potential. As the neutral scalar $\phi^0$ mixes with the SM Higgs, the Higgs searches~\cite{Chatrchyan:2012tx, Aad:2012tfa, Chatrchyan:2012ufa} apply. On the other hand the charged components decay to $W$-bosons and SM Higgs bosons. However the current searches for a singly-charged scalar~\cite{Aad:2012rjx, Aad:2012tj,
Chatrchyan:2012vca, Aaltonen:2011aj, Aaltonen:2009ke, Abazov:2009wy, Abazov:2009aa, Abazov:2009ae,Abazov:2008rn} and a doubly-charged scalar~\cite{Chatrchyan:2012ya,ATLAS:2012hi} do not apply, because both searches assume a coupling to SM fermions. 
Triply-charged scalars have also been briefly studied~\cite{Babu:2009aq} and a rough bound of $\sim 120$ GeV
has been estimated for triply-charged scalars with displaced decay vertices based on the D0 and CDF searches for long-lived massive particles. 

A proper collider study relies on a detailed study of the mass spectrum and the different decay channels, which is beyond the scope of this short summary.

\section{Detailed Study of a Specific Model}
\label{sec:example}

\subsection{Model}
In order to demonstrate the LHC reach with regard to minimal UV completions of D7 $\Delta L=2$ operators, we study a model with a scalar leptoquark $\phi$ and a vector-like quark $\chi$ with quantum numbers
\begin{equation}
\phi\sim\left(\bar 3,1,\frac13\right),\qquad \chi\sim\left(3,2,-\frac56\right)\;.
\end{equation}
These particles arise in the minimal UV completions of $\mathcal{O}_3=LLQ\bar{d}H$ and 
$\mathcal{O}_8=L\bar{d}\bar{e}^\dagger\bar{u}^\dagger H$ operators, for which more details are available in Appendix~\ref{app:uvcompletion}.
The Yukawa couplings and bare mass terms of the new exotic particles are given by
\begin{align}
-\mathcal{L}&=\mu^2_\phi \phi^\dagger\phi + m_\chi \bar\chi\chi
+\left(Y^{LQ\phi}_{ij} L_i Q_j \phi
+Y^{L\bar\chi\phi}_{i} L_i\bar\chi \phi^\dagger 
+ Y^{\bar d\chi H}_{ij} \bar d_i \chi_j H +\hc\right) \\\nonumber
&+\left(Y^{\bar e\bar u \phi}_{ij} \bar e_i \bar u_j \phi^\dagger+\hc\right) .
\end{align}
Besides the SM gauge symmetry group, we have to demand baryon-number conservation, in order to forbid the operators $Y^{QQ\phi}_{ij} Q_iQ_j\phi^\dagger$ and $Y^{\bar d\bar u\phi}_{ij} \bar d_i \bar u_j \phi$, which induce proton decay in analogy to Ref.~\cite{Angel:2013hla}.

\subsection{Neutrino Mass Generation}
\label{subsec:numassgen}

\begin{figure}[tp]
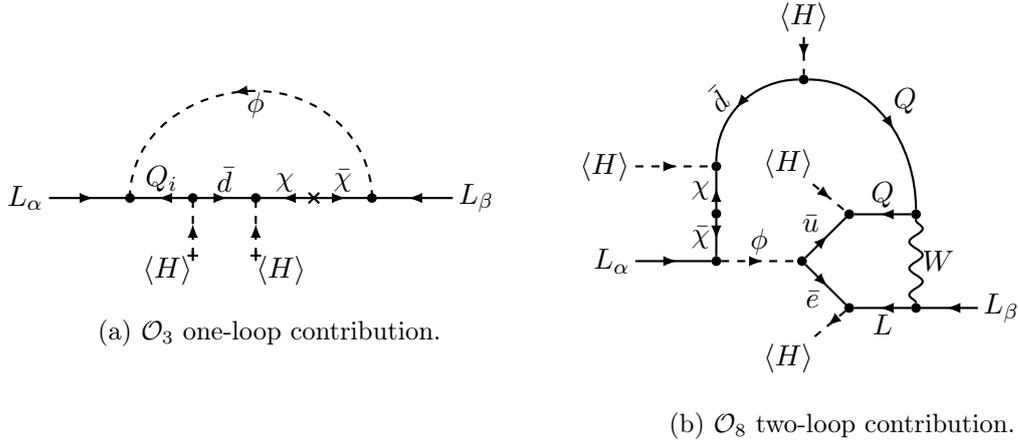
\centering
\begin{subfigure}{0.49\linewidth}
\FMDG{NuMass}
\caption{$\mathcal{O}_3$ one-loop contribution.}
\label{fig:NuMass}
\end{subfigure}
\begin{subfigure}{0.49\linewidth}
\FMDG{NuMass2}
\caption{$\mathcal{O}_8$ two-loop contribution.}
\label{fig:NuMassO8}
\end{subfigure}
\caption{Neutrino mass diagrams.}
\label{fig:NuMassAll}
\end{figure}

In this model, neutrino mass receives its dominant contribution from the radiative diagram of \Figref{fig:NuMass}. The two-loop $\mathcal{O}_8$ contribution depicted in \Figref{fig:NuMassO8} as well as the corresponding three-loop contribution, which is obtained from \Figref{fig:NuMassO8} by connecting two external Higgs lines, are generally subdominant unless the coupling of the leptoquark $\phi$ to RH fermions is much larger, $|Y^{\bar e \bar u \phi}_{i3}| \gg |Y^{LQ\phi}_{j3}|$. The neutrino mass matrix is proportional to the down-type quark mass matrix, and it is dominated by the bottom quark. For simplicity we will assume that the vector-like quark only mixes with the third generation quarks and set all couplings to the first two generation quarks to zero. In addition we will focus on the $\mathcal{O}_3$ contribution, neglect the $\mathcal{O}_8$ contributions and assume $Y^{\bar e \bar u \phi}_{ij}=0$.
Decomposing the vector-like quark $\chi$ and $\bar\chi$ into its components with respect to SU(2)$_L$, we write
\begin{align}
\chi&=\begin{pmatrix}B^\prime\\Y\end{pmatrix} , &
\bar\chi&=\begin{pmatrix}\bar Y\\\bar B^\prime\end{pmatrix}.
\end{align}
$\bar Y$ and $Y$ form a Dirac pair with mass $m_Y=m_\chi$ and $\bar B^\prime$ and $B^\prime$ mix with the gauge eigenstate of the bottom quark $b^\prime$,
\begin{align}
&\left(\begin{array}{c}
\bar{b} \\ \bar{B}
\end{array}\right) = 
\left(\begin{array}{cc} c_1 & s_1 \\ -s_1 & c_1\end{array}\right)^\dagger
\left(
\begin{array}{c}
\bar{b}^\prime \\ \bar B^\prime
\end{array}
\right),
& \left(
\begin{array}{c}
b \\ B
\end{array}
\right) = 
\left(\begin{array}{cc} c_2 & s_2 \\ -s_2 & c_2\end{array}\right)^\dagger
\left(
\begin{array}{c}
b^\prime \\ B^\prime
\end{array}
\right),
\end{align}
forming the mass eigenstates $b$ and $B$. The physical masses are
\begin{align}
m_b^2 &= m_{b^\prime}^2-m_{bB}^2 \frac{m_\chi^2}{m_\chi^2-m_{b^\prime}^2} \;, &
m_B^2 &= m_\chi^2 + m_{bB}^2 \frac{m_{b^\prime}^2}{m_\chi^2-m_{b^\prime}^2}
\end{align}
with $m_{bB}=Y_3^{\bar d\chi H} v /\sqrt{2}$, $m_{b^\prime}=y_b v/\sqrt{2}$ and the mixing angles are given by
\begin{align}
& s_1=\frac{m_{bB} \ m_\chi}{m_\chi^2-m_{b^\prime}^2},
& s_2=\frac{m_{bB} \ m_{b^\prime}}{m_\chi^2-m_{b^\prime}^2} 
\end{align}
with $c_{1,2}=\sqrt{1-s_{1,2}^2}$.
After electroweak symmetry breaking, we calculate the radiatively generated neutrino mass as
\begin{equation}
(m_\nu)_{ij}= \frac{3}{16\pi^2} \left(Y_{i3}^{LQ\phi} Y_j^{L\bar\chi\phi} +(i\leftrightarrow j) \right) m_{bB}\frac{m_b m_B}{m_B^2-m_b^2}\left(\frac{m_B^2\ln \frac{m_B^2}{m_\phi^2}}{m_\phi^2-m_B^2}-\frac{m_b^2\ln\frac{m_b^2}{m_\phi^2}}{m_\phi^2-m_b^2}\right)\;.
\label{eq:numasso3b1a}
\end{equation}
In the limit $m_b\ll m_B,m_\phi$ this reduces to 
\begin{equation}
(m_\nu)_{ij} = \frac{3}{16\pi^2} \left(Y_{i3}^{LQ\phi} Y_j^{L\bar\chi\phi} +(i\leftrightarrow j)\right) m_{bB}
\frac{m_b m_B}{m_\phi^2-m_B^2}\ln \frac{m_B^2}{m_\phi^2}\;.
\end{equation}
Thus there is one almost massless neutrino and two massive neutrinos. 

Next we would like to use the low-energy parameters (the PMNS matrix as well as the neutrino masses) to determine the Yukawa couplings in terms of the high-scale parameters. 
The flavour structure of the neutrino mass matrix can be parameterised by vectors $a_\pm$ and a common factor $\alpha$, 
\begin{equation}\label{eq:NuMassStruct}
m_\nu = \alpha (a_+ a_-^T + a_- a_+^T)\;,
\end{equation}
i.e.\ the neutrino mass matrix is generated by multiplying two different vectors $a_\pm$ symmetrically. On the other hand it can be written in terms of the low-energy parameters for normal (NO) as well as inverted (IO) mass ordering,
\begin{align}\label{eq:mNuLowEnergy}
m_\nu^{NO} &= m_2 u_2^* u_2^\dagger + m_3 u_3^* u_3^\dagger \ ,&
m_\nu^{IO} &= m_1 u_1^* u_1^\dagger + m_2 u_2^* u_2^\dagger \ ,
\end{align}
where $m_i$ are the neutrino masses and $U=(u_1,u_2,u_3)$ is the PMNS matrix.
We can rewrite the right-most expression of \Eqref{eq:NuMassStruct} as
\begin{equation}
\alpha (a_+a_-^T + a_- a_+^T)=\frac{\alpha}{2} \left[ \left( \frac{a_+}{\zeta}+\zeta a_-\right)\left( \frac{a_+}{\zeta}+\zeta a_-\right)^T -\left( \frac{a_+}{\zeta}-\zeta a_-\right)\left( \frac{a_+}{\zeta}-\zeta a_-\right)^T \right]
\end{equation}
and match it onto \Eqref{eq:mNuLowEnergy} to obtain the vectors $a_\pm$ in terms of the low-energy parameters:
\begin{align}
a_{\pm}^\mathrm{NO}&=\frac{\zeta^{\pm 1}}{\sqrt{2\alpha}} \left(\sqrt{m_2} u_2^* \pm i \sqrt{m_3} u_3^*\right) ,&
a_{\pm}^\mathrm{IO}&=\frac{\zeta^{\pm 1}}{\sqrt{2\alpha}} \left(\sqrt{m_1} u_1^* \pm i \sqrt{m_2} u_2^*\right)\;.
\end{align}
The complex parameter $\zeta$ is a free parameter not determined by low-energy physics.

We use the best fit values (v1.2) of the NuFIT collaboration~\cite{GonzalezGarcia:2012sz}\footnote{The newest best fit values in v1.3 of the NuFIT collaboration are slightly changed. See~\cite{Tortola:2012te, Fogli:2012ua} for other global fits to the neutrino oscillation data.} assuming normal ordering:
\begin{align}
\sin^2\theta_{12}&=0.306\,, &
\Delta m_{21}^2&=7.45\times 10^{-5} \eV^2\,,\nonumber\\\label{eq:nuParam}
\sin^2\theta_{13}&=0.0231\,, &
\Delta m_{31}^2& =2.417\times 10^{-3}\eV^2 \,,\\\nonumber
\sin^2\theta_{23}&=0.446\;.
\end{align}
Furthermore we set the lightest neutrino mass to zero and assume vanishing CP phases in the PMNS matrix, i.e.\ $\delta=\varphi_1=\varphi_2=0$.

\subsection{Constraints From Flavour Physics and Neutrino-less Double-Beta Decay}

Experimental constraints on flavour violating processes already constrain the parameter space. Similarly to the two-loop model in Ref.~\cite{Angel:2013hla}, we expect the most stringent constraints from lepton-flavour violating processes, in particular from the $\mu \rightarrow e$ transition. We calculated $\mu\to e \gamma$, $\mu\to eee$ as well as $\mu N \to e N$ conversion in nuclei and compared the results with the current experimental limits. We use the contributions calculated in Ref.~\cite{Angel:2013hla} and add the contributions from the additional coupling of the leptoquark to the vector-like lepton. The Wilson coefficients are included in \Appref{app:LFV}.

\begin{figure}[tp]\centering
\begin{subfigure}{0.49\linewidth}
\includegraphics[width=\linewidth]{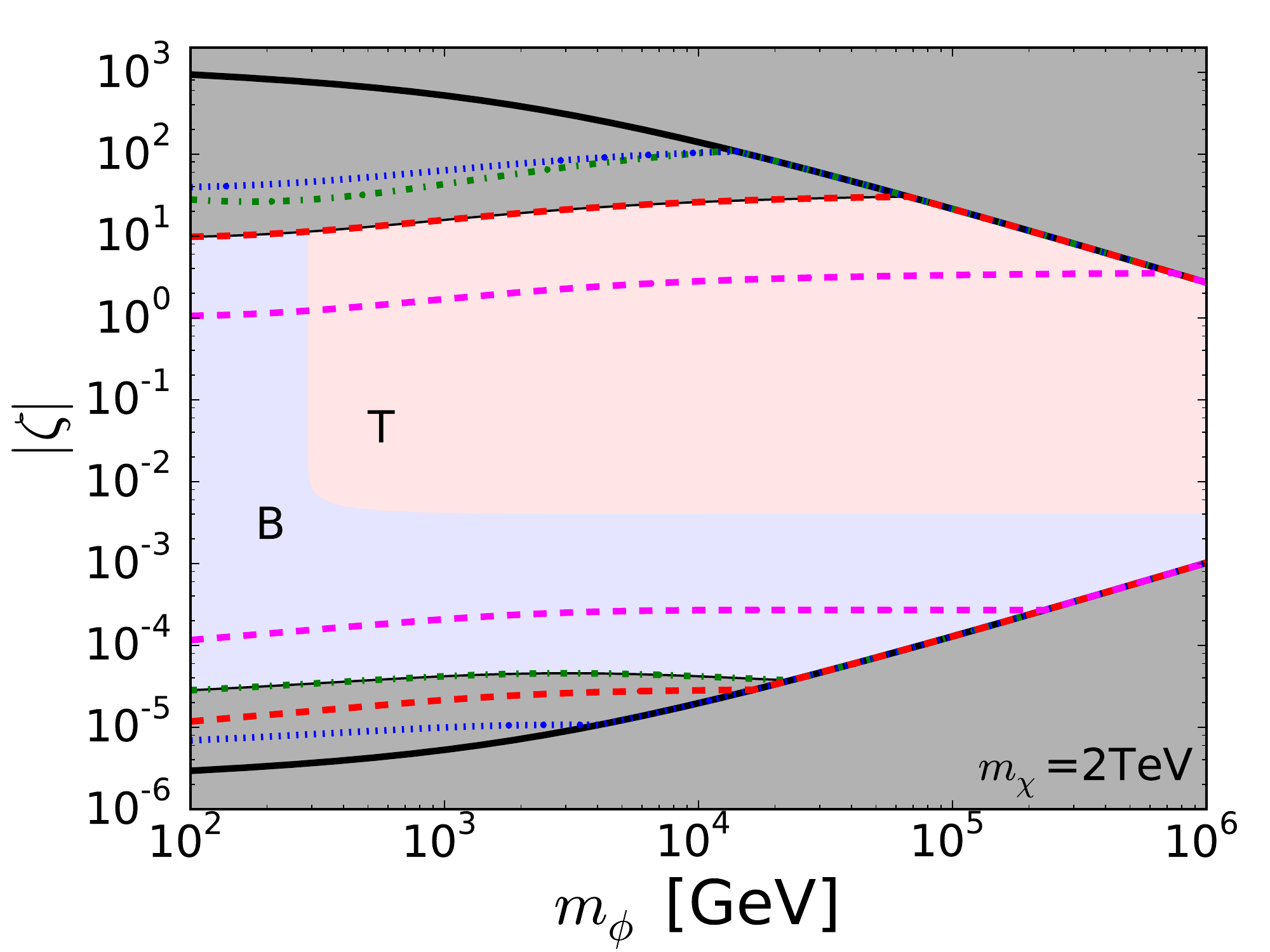}
\end{subfigure}
\hfill
\begin{subfigure}{0.49\linewidth}
\includegraphics[width=\linewidth]{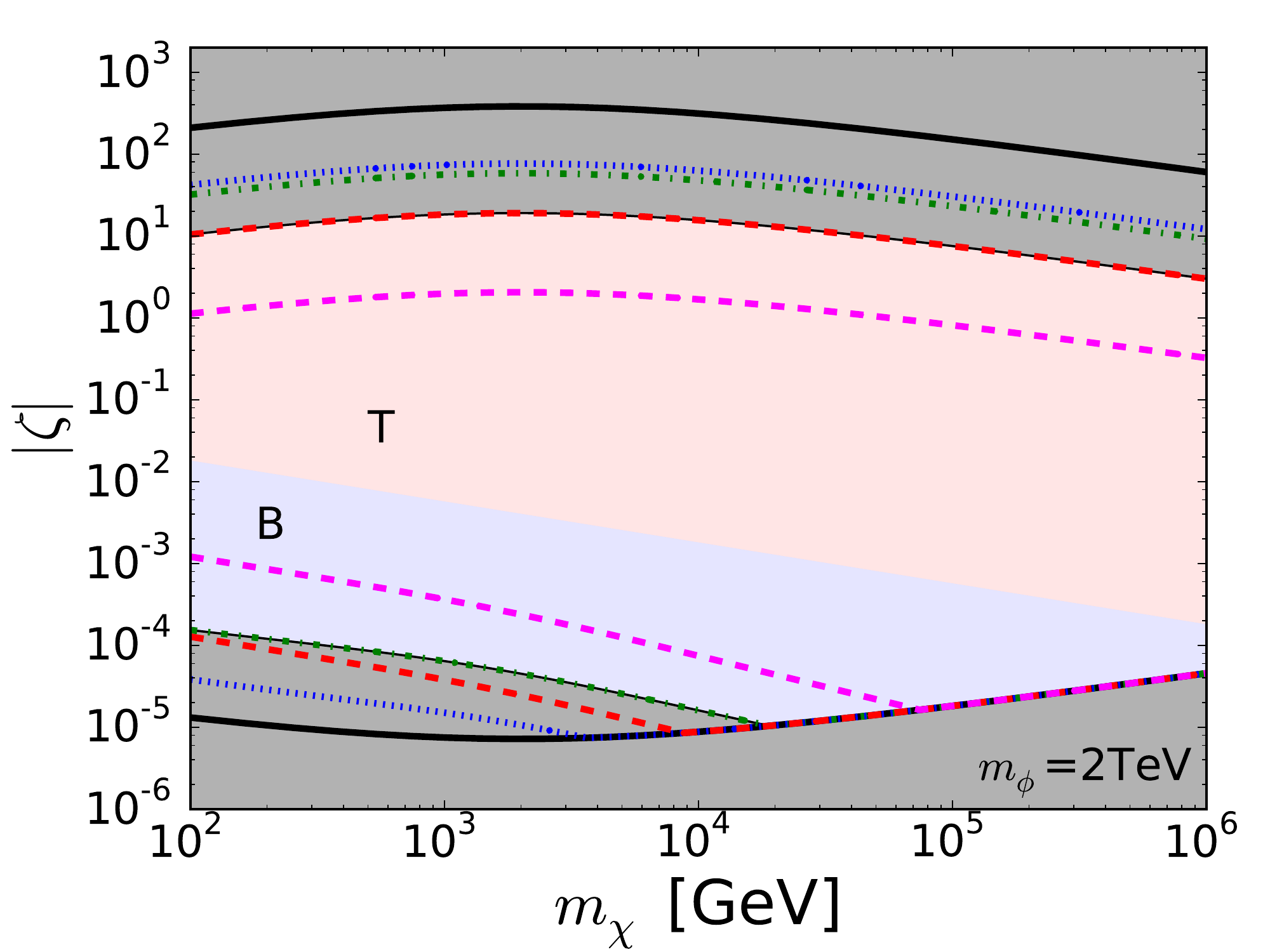}
\end{subfigure}
\caption{Constraints on $\zeta$ and the two different experimental search regions for the leptoquark $\phi$. The grey shaded region is excluded. The light blue shaded region (B) indicates the allowed region with Br$(\phi\to b\nu)\approx 100\%$. The light red shaded region (T) has Br$(\phi\to b\nu)<100\%$. 
The solid black lines indicate the bound from perturbativity of Yukawa couplings. We require $\max(|Y^{LQ\phi}_{ij}|,|Y^{L\bar\chi\phi}_{ij}|)<1$. The green dot-dashed, blue dotted, red dashed lines show the limits Br$(\mu\to e \gamma)< 5.7\times10^{-13}$ \cite{Adam:2013mnn}, Br$(\mu\to eee)<10^{-12}$ \cite{PDG:2012}, and Br$(\mu \mathrm{Au}\to e\mathrm{Au})<7\times 10^{-13}$ \cite{PDG:2012}. The magenta dashed line indicates the projected experimental sensitivity of $10^{-16}$ to measure $\mu\mathrm{Ti}\to e\mathrm{Ti}$ conversion in titanium in Mu2E at FNAL and COMET at J-PARC~\cite{Carey:2008zz, Kutschke:2011ux, Cui:2009zz}.}
\label{fig:zetaConstraints}
\end{figure}

As all parameters are fixed by the leptonic mixing and the neutrino masses, the constraints directly translate to a constraint on the complex rescaling parameter $\zeta$, more precisely on $|\zeta|$.  The phase of $\zeta$ drops out in the flavour physics amplitudes, at least for the leading contributions, because they are of the form $Y_i^{L\bar\chi\phi*}Y_j^{L\bar\chi\phi}$ and $Y_i^{LQ\phi*}Y_j^{LQ\phi}$. We present the constraints on $|\zeta|$ while varying one of the masses $m_{\phi,\chi}$ in \Figref{fig:zetaConstraints}. The other mass is fixed to 2 TeV. The grey shaded region is excluded (see the caption for an explanation of the different exclusion lines). Our main result is that within the bounds on $|\zeta|$ from LFV experiments there are two regions, separated by a sharp transition, with very different search strategies for the leptoquark $\phi$.
The light blue shaded region (region B) indicates the allowed region with Br$(\phi\to b\nu)\approx 100\%$. The light red shaded region (region T) has Br$(\phi\to b\nu)<100\%$. We discuss both of these regions in \Secref{sec:detail:leptoquark}.

In addition to constraints from flavor violating processes, there are constraints from lepton-number violating processes, like neutrino-less double beta decay. The relevance of neutrino-less double beta decay for radiative neutrino mass models with coloured particles in the loop has been illustrated in Ref.~\cite{Choubey:2012ux}. More generally, Ref.~\cite{Bonnet:2012kh} studied possible contributions to neutrino-less double beta decay  by systematically decomposing the dimension-9 operator.

This specific model will lead to additional short-range contributions to neutrinoless double beta decay via couplings to the first generation of quarks. As neutrino mass does not depend on the coupling to the first generation of quarks, this bound can always be satisfied by setting these couplings to zero without affecting the mechanism of neutrino mass generation.
This leaves the long-range contribution via an exchange of active neutrinos, which is controlled by the effective mass
\begin{equation}
\langle m_{ee}\rangle=\sum U_{ei}^2m_i\ .
\end{equation}
As the minimal framework leads to a strong mass hierarchy, there are currently no competitive constraints from neutrino-less double beta decay, similarly to the discussion in Ref.~\cite{Angel:2013hla}.

\subsection{Vector-Like Quark Search}
\label{sec:example:vlq}
\begin{figure}[tp]
\centering
\includegraphics[width=0.5\linewidth]{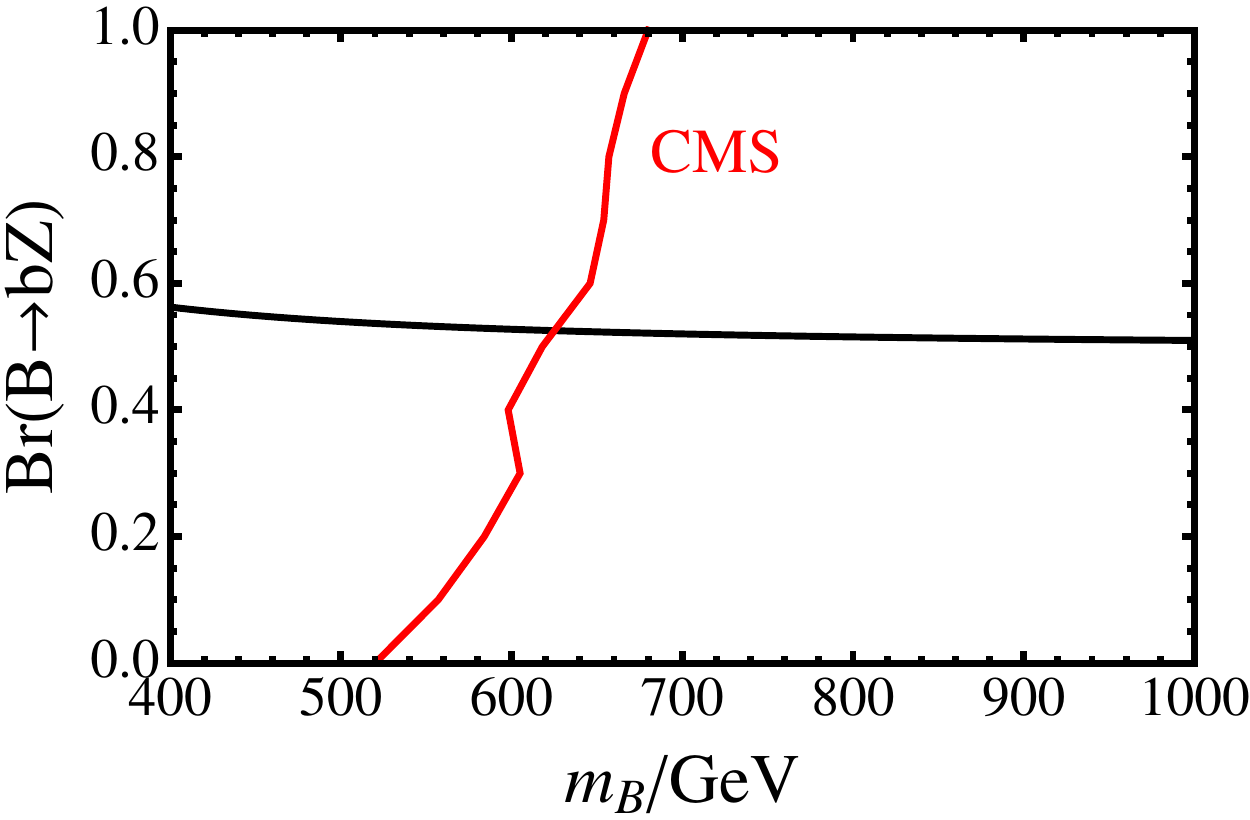}
\caption{The branching ratio of $B\rightarrow bZ$ as a function of the heavy $B$ mass with the observed limit from CMS shown in red. }
\label{fig:cmsheavyquarks}
\end{figure}

As discussed in Sec.~\ref{subsec:vectorlikequarks}, the mass eigenstate $B$ will decay mainly through 
$B\rightarrow Zb$ and $B\rightarrow Hb$ while the third channel $B\rightarrow W^- t$ is highly suppressed
due to the small mixing between the heavy vector-like quark $B$ and the SM $b$-quark.
The dominant branching ratios obey the relation
\begin{eqnarray}
\frac{ {\rm Br} (B\rightarrow Z b)}{{\rm Br}(B\rightarrow H b)}&=&
\frac{\lambda(1,r_b,r_Z)^{1/2}}{\lambda(1,r_b,r_H)^{1/2}}
 \frac{ 1+r_Z^2-2 r_b^2-2 r_Z^4+r_b^4+r_Z^2 r_b^2}{1+6 r_b^2-r_H^2+r_b^4-r_b^2r_H^2} \ ,
\end{eqnarray}  
where $r_{b,H,Z}=m_{b,H,Z}/m_B$ and 
\begin{align}\label{eq:defLambda}
\lambda(M,m_1,m_2) &=M^4+m_1^4+m_2^4-2 M^2 m_1^2-2 M^2 m_2^2-2 m_1^2m_2^2\\
&=(M^2-(m_1+m_2)^2)(M^2-(m_1-m_2)^2)\;. \nonumber
\end{align}
We can easily read our limit on the mass of $B$, $m_B\gtrsim 620 \ \rm{GeV}$, from the dedicated CMS search as a function of the branching ratios in 
Fig.~\ref{fig:cmsheavyquarks}.

\subsection{Leptoquark Searches}
\label{sec:detail:leptoquark}

In the following subsection we take $L\equiv \{e,\mu,\tau\}$ and $l\equiv \{e,\mu\}$. 
The scalar leptoquark $\phi$ can be pair-produced at the LHC via $gg$ fusion and $q\bar{q}$ annihilation. The cross section $\sigma(pp\to\phi\phi)$ is determined purely by colour charge and therefore depends only on the mass $m_\phi$. We use NLO \textsc{Prospino2} \cite{Kramer2004df} cross sections for the LHC running at $\sqrt{s}=8$~TeV, which gives $\sigma(pp\to \phi\phi)=82$ (23.5) fb for $m_\phi=500$ (600) GeV. We ignore the t-channel lepton exchange contribution and single production $qg\to \phi L$, since these will be suppressed by powers of small Yukawa couplings.

Upon pair production, the leptoquarks will decay with branching ratios dependent on the parameters $Y_{L3}^{LQ\phi}$ and $Y_3^{\bar{d} \chi H}$ relevant to neutrino mass generation.
The partial decay widths are
\begin{align}
 \Gamma(\phi\rightarrow  L t)&=\frac{m_\phi}{8\pi}\left| Y_{L3}^{LQ\phi}\right|^2 f(m_\phi,m_{L},m_t) \ , \\
 \Gamma(\phi\rightarrow  \nu_L  b )&=\frac{m_\phi}{8\pi}\left(\left|Y_{L3}^{LQ\phi} c_2\right|^2 + \left|Y_{L}^{L\bar\chi\phi} s_1 \right|^2\right) f(m_\phi,m_{\nu_L},m_{ b}) \label{eq:phi2nub} \\\nonumber
 &-\frac{m_\phi}{4\pi}\mathrm{Re}\left(Y_{L3}^{LQ\phi} c_2  Y_{L}^{L\bar\chi\phi} s_1^* \right) f^\prime(m_\phi,m_{\nu_L},m_{ b}) \ , \\
\Gamma(\phi\rightarrow  \nu_L B )&=\frac{m_\phi}{8\pi}\left(\left|Y_{L3}^{LQ\phi} s_2\right|^2 + \left| Y_{L}^{L\bar\chi\phi} c_1 \right|^2\right) f(m_\phi,m_{\nu_L},m_{B}) \label{eq:phi2nuB} \\\nonumber
& + \frac{m_\phi}{4\pi} \mathrm{Re}\left(Y_{L3}^{LQ\phi} s_2 Y_{L}^{L\bar\chi\phi} c_1^* \right) f^\prime(m_\phi,m_{\nu_L},m_{B}) \ , \\
 \Gamma(\phi\rightarrow L Y)&= \frac{m_\phi}{8\pi} \left| Y_{L}^{L\bar\chi\phi}\right|^2 f(m_\phi,m_{L},m_{Y}) \ ,
\end{align}
where $b$, $B$  are the two heaviest down-type quark mass eigenstates and the functions $f$, $f^\prime$ are defined as
\begin{align}
f(M,m_1,m_2)&=\frac{\left( M^2-m_1^2-m_2^2\right) \lambda(M,m_1,m_2)^{1/2} }{2\,M^4} \ ,\\
f^\prime(M,m_1,m_2)&=\frac{m_1 m_2 \lambda(M,m_1,m_2)^{1/2} }{M^4} 
\end{align}
with $\lambda$ given in \Eqref{eq:defLambda}. 
The term in the second lines of \Eqref{eq:phi2nub} and \Eqref{eq:phi2nuB} is neglible because it is suppressed by the neutrino mass. Note that the phase of $\zeta$ drops out in all decay widths.
Nonzero couplings that are not constrained by the neutrino mass generation generally open extra decay channels.
Since we are only interested in the consequences of neutrino mass generation, all these couplings are taken to be zero.

In the following we will concentrate on the region in parameter space with $m_Y,m_B\gg m_\phi$: each leptoquark may decay into either $b\nu$ or $tL$, resulting in $b\nu b\nu$, $b\nu tL$ or $tLtL$ after pair production. The branching ratios are determined by the single complex parameter $\zeta$ after fitting to low energy parameters as described in Sec.~\ref{subsec:numassgen}. 
There are two regions of interest:
\begin{itemize}
\item Region B where the branching ratio Br$(\phi\to b\nu)\approx 100\%$, either because the other channels are kinematically not accessible for $m_\phi\lesssim m_t$ or $\left|Y^{LQ\phi}\right|\ll \left|Y^{L\bar\chi\phi}\right|$. It is shaded light blue in \Figref{fig:zetaConstraints}.
\item Region T in which all decay channels are open. It is shaded light red in \Figref{fig:zetaConstraints}.
\end{itemize}

\begin{figure}[t]\centering
 \includegraphics[width=8cm]{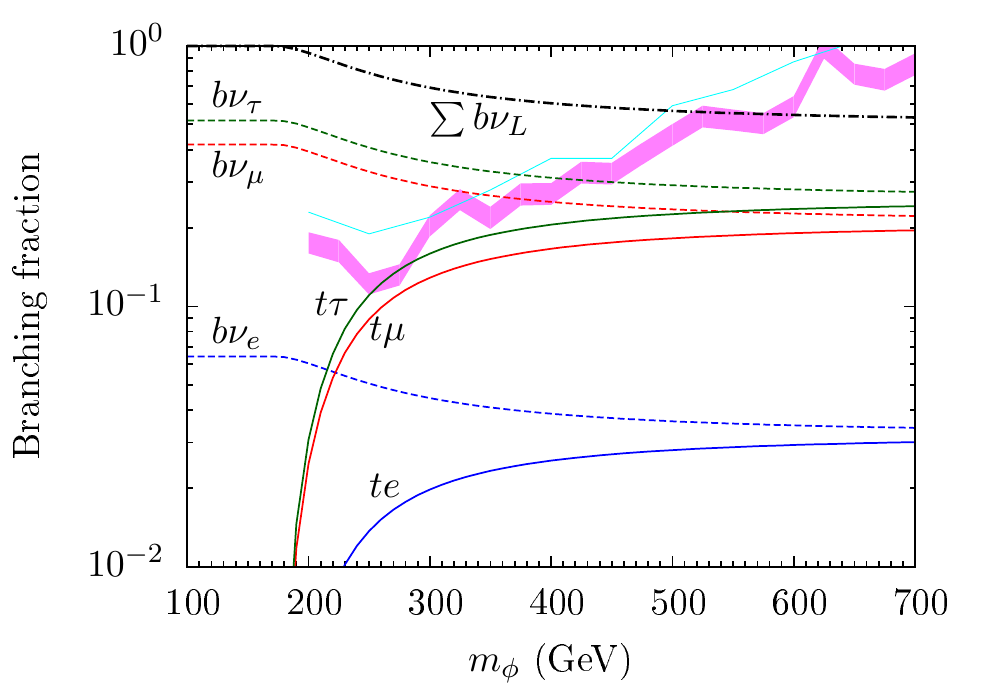}

	 \caption{Branching fractions for $\phi$ as a function of $m_\phi$ in the region T. Also shown are limits on $\mathrm{Br}(\phi\to \sum_L b\nu_L)$ from sbottom pair searches of ATLAS (light blue) and CMS (the limit line lies somewhere within the magenta band).}
 \label{figbranch1}
\end{figure}

In region B we have $\mathrm{Br}(\phi\to \sum b\nu_L)\approx 1$, resulting in a $bb\slashed{E}_T$ final state for which sbottom pair searches can be directly applied \cite{Aad2013ija,CMS-PAS-SUS-13-018}. In this case $m_\phi$ is constrained to be $\gtrsim 730$~GeV at 95\% CL. Fig.~\ref{figbranch1} shows branching ratios for region T in the case of normal ordering. The hierarchy between Br$(\phi\to t \mu$) $\approx$ Br$(\phi\to t \tau$) and Br$(\phi\to t e$) is larger for normal compared to inverted mass ordering.\footnote{For example, at $m_\phi=500$~GeV, normal ordering gives $(0.028,0.183,0.226)$ for $\mathrm{Br}(\phi\to te,t\mu,t\tau)$, whilst inverted ordering gives $(0.070,0.165,0.202)$.} Hence there will be slightly more electrons in final states for the inverted mass ordering. The relative size of Br$(\phi\to t \mu$) and Br$(\phi\to t\tau$) is controlled by the atmospheric mixing angle $\theta_{23}$, i.e. for $\theta_{23}> \pi/4$, Br$(\phi\to t\mu)>\mathrm{Br}(\phi\to t\tau)$ and we expect the 
limits to get slightly stronger. In the limit of large $m_\phi$ it is apparent that $\mathrm{Br}(\phi\to \sum b\nu_L)\approx \mathrm{Br}(\phi\to \sum tL) \approx 0.5$. 

In region T we can now calculate the branching fractions to LHC-reconstructable final states.\footnote{We do not attempt to reconstruct $\tau$ leptons since this will not improve sensitivity. CMS has performed a dedicated search for leptoquarks decaying to $t\tau$ \cite{CMS-PAS-EXO-12-030}; the resulting bounds are not competitive with the bounds found henceforth.} The most frequent final state is $bb\slashed{E}_T$ at about $30\%$; as we will see, because $\mathrm{Br}(\phi\to \sum b\nu_L)$ is always greater than $50\%$, existing sbottom pair searches alone can provide a bound of $m_\phi > 500$~GeV. But can another final state compete? The next most frequent final state is $l bbjj\slashed{E}_T$ at about $22\%$; in this case, searches for top squark pairs in final states with one isolated lepton are applicable.
About $8\%$ of the time a two-lepton final state is produced; again, searches for top squark pairs are applicable.
Three- and four-lepton final states are also predicted by this model in $\lesssim 1\%$ of events.
For $m_\phi = 600$~GeV, where we will find the existing bound lies, one expects $\approx500$ leptoquark pair events in the $\sqrt{s}=8$~TeV dataset. When compared to existing limits, it turns out there are simply not enough three- or four-lepton events to provide a competitive limit \cite{Aad2014pda,Aad2014iza}. However it is possible that, with more data at $\sqrt{s}=13$~TeV, these final states can be competitive.

In the following subsections we will cover three final states of interest, namely $bb\slashed{E}_T$, $l\slashed{E}_T + (b\text{-})jets$, and $l^+l'^-\slashed{E}_T + jets$. Our aim is to recast LHC stop searches \cite{ATLAS-CONF-2013-037,Aad2014qaa} in order to constrain $m_\phi$.

\subsubsection{Event samples and reconstruction}

We generated two hadron-level signal samples at $m_\phi=(500,600)$~GeV\footnote{We also used $m_\chi=2$~TeV and $s_1=0.01$, but the branching ratios do not depend on the choice of $m_\chi$ and $s_1$ as long as $m_\chi> m_\phi$.} using \textsc{Pythia 8.180} with default tune \cite{Sjostrand2006za, Sjostrand2007gs}; each contained $5\times10^6$ pair-produced leptoquark events where at least one leptoquark decays to $tL$. A validation set of $10^7$ $t\bar{t}$ events where at least one $t$ decays leptonically was also generated using \textsc{Pythia}, normalised to the predicted NNLO+NNLL cross section of $235\times[1-\mathrm{Br}(W\to \mathrm{hadrons})^2]$~pb = 137~pb \cite{Cacciari2011hy, Baernreuther2012ws, Czakon2012zr, Czakon2012pz, Czakon2013goa, Czakon2011xx}. 
Lastly we used \textsc{MadGraph5 v1.5.10} and \textsc{Pythia} to generate a validation set of $10^5$ stop pair events, where the stops each decayed to a top and neutralino, $\tilde{t}_1 \to t\tilde{\chi}_1^0$; we took $m(\tilde{t}_1,\tilde{\chi}_1^0)=(600,50)$~GeV.

The event samples were reconstructed after passing through the \textsc{Delphes 3.0.12} detector fast-simulation \cite{deFavereau2013fsa}, both with and without simulated pileup. Jets were reconstructed with \textsc{FastJet 3.0.6} \cite{Cacciari2011ma} using the anti-$k_t$ clustering algorithm \cite{Cacciari2008gp} with radius parameter 0.4, and were required to have $p_T>20$~GeV. We used a flat $b$-tag rate of 70\%, with a rejection factor of 5 (140) for jets initiated by charm (light) quarks. Electrons were considered isolated if $\sum p_T$, the scalar sum of the $p_T$ of inner detector tracks with $p_T>1$~GeV within a $\Delta R=0.2$ cone surrounding the electron candidate, was less than $10\%$ of the electron $p_T$. Muons were considered isolated if $\sum p_T$, defined as above, was less than 1.8~GeV. Otherwise, the default \textsc{Delphes} ATLAS card was used. 
In the simulations with pileup we used a mean pileup $\mu=21$, and pileup subtraction was performed using default parameters; the neutral pileup subtraction uses the jet area method \cite{Cacciari2007fd, Cacciari2008gn} with average contamination density $\rho$ calculated 
using a 
$k_t$ jet clustering algorithm with radius parameter 0.6. We note that this pileup subtraction method does not match that used in either of the ATLAS analyses. The results simulated with pileup therefore serve only as an indicator of pileup effects.

Further cuts were made with the aid of the \textsc{MadAnalysis5} v1.1.10beta \textsc{SampleAnalyzer} framework \cite{Conte2012fm}. For preselection we required isolated leptons and
\begin{align}
 |\eta_e| <2.47, && |\eta_\mu| < 2.4, && |\eta_j| <2.5, && p_T^l > 10 \text{ GeV}.
\end{align}
We rejected jets within $\Delta R=0.2$ of a preselected electron, and leptons within $\Delta R=0.4$ of remaining jets. 

Each of the stop search analyses use variants of $m_{T2}$, known as the Cambridge $m_{T2}$ or stransverse mass variable \cite{Lester1999tx, Barr2003rg}, as a powerful discriminant of signal over background. For events where mother particles are pair produced and subsequently decay to two visible branches along with invisible momentum, such as in leptonic or semi-leptonic $t\bar{t}$ decays, $m_{T2}$ can be constructed to have an upper limit at the mother particle mass. It is defined as
\begin{align}
 m_{T2}&(\vec{p}_T^i,\vec{p}_T^j,\slashed{\vec{p}}_T) =
 \min\limits_{\vec{u}_T+\vec{v}_T=\slashed{\vec{p}}_T}
 \left\{ \max\left[ m_T(\vec{p}_T^i,\vec{u}_T),m_T(\vec{p}_T^j,\vec{v}_T) \right] \right\},
\end{align}
where $\slashed{\vec{p}}_T$ is the missing transverse momentum, $\vec{p}_T^i$ and $\vec{p}_T^j$ are the transverse momenta of two visible decay branches, and $m_T$ is the usual transverse mass calculated assuming some mass for the invisible particle associated with that branch. It can be thought of as the minimum mother particle mass consistent with pair production, the decay hypothesis, and the observed kinematics. We calculated $m_{T2}$ using the publicly available bisection method codes of Refs.~\cite{Cheng2008hk,Bai2012gs}.

\subsubsection{$bb\slashed{E}_T$}

The $bb\slashed{E}_T$ final state arises primarily from the decay $\phi\phi \to b\nu b\nu$. There are also contributions from the other decay chains, where either leptons are missed or hadronically decaying taus are produced; these contributions will be subleading and additive, and will generally appear with extra hard jet activity in the event which may be vetoed in analyses. We will ignore them to obtain a slightly conservative limit. 

Constraints on the production cross section of sbottom pairs decaying via $\tilde{b}_1\to b\tilde{\chi}_1^0$ have been provided by both ATLAS and CMS \cite{Aad2013ija,CMS-PAS-SUS-13-018}. Along the contour $m_{\tilde{\chi}_1^0}=0$, this provides a limit on the production cross section $\sigma(pp \to \phi\phi) \times \mathrm{Br}(\phi \to b\nu)^2$, and therefore on $\mathrm{Br}(\phi \to b\nu)$. These limits are reproduced in Fig.~\ref{figbranch1}.\footnote{The ATLAS limit on $\mathrm{Br}(\phi \to b\nu)$ can be read off the auxiliary Figure~5. The CMS limit on $\sigma(pp \to \phi\phi) \times \mathrm{Br}(\phi \to b\nu)^2$ can be read off Figure~6 and converted to a limit on $\mathrm{Br}(\phi \to b\nu)$ using the NLO value of $\sigma(pp \to \phi\phi)$ from \textsc{Prospino2}.} The existing 95\% CL limit from the CMS search for region T is somewhere between $m_\phi >$~520--600~GeV.

\subsubsection{$l\slashed{E}_T + (b\text{-})jets$}

The single lepton final state is produced primarily through the mixed decay $\phi\phi \to b\nu tL \to l jj bb \slashed{E}_T$, where the top decays hadronically. 
This final state is the same as for semi-leptonically decaying top pairs, which is the primary SM background. It can also be given by stop pairs decaying via the chains $\tilde{t}_1 \to b\tilde{\chi}_1^{\pm}\to b W^{(*)}\tilde{\chi}_1^0 \to b l\nu\tilde{\chi}_1^0$ or $\tilde{t}_1 \to t^{(*)}\tilde{\chi}_1^0 \to bl\nu\tilde{\chi}_1^0$. ATLAS and CMS have performed searches for stop pairs in the single lepton final state, with no significant excess observed \cite{ATLAS-CONF-2013-037,Chatrchyan2013xna}.\footnote{In the time since this analysis was performed, ATLAS submitted a more detailed search in this channel \cite{Aad2014kra}.} 
In this section we will recast the ATLAS analysis.

\begin{table*}\centering \footnotesize
 \begin{tabular}{ccccccc} 
\toprule
	&	& SRtN2		& SRtN3 	& SRbC1		& SRbC2 	& SRbC3 	\\ \midrule
	$m_{\tilde{t}_1}=600$~GeV 
	& $\mathcal{A}\varepsilon$ ATLAS (\%)
		& 2.7		& 2.3		& 5.7		& 1.7		& 0.84		\\
		$m_{\tilde{\chi}_1^0}=50$~GeV 
		& $\mathcal{A}\varepsilon$  obtained (\%)
		& 2.0 (2.1) 	& 1.4 (1.5)	& 5.8 (5.6)	& 1.8 (1.6)	& 1.0 (0.83) 	\\ \midrule
  $m_\phi=500$~GeV
	& $N$
		& 21 (22)	& 14 (14)	& 75 (74)	& 28 (26)	& 16 (14)	\\ 
  $m_\phi=600$~GeV 
	& $N$
		& 7.8 (8.3) 	& 5.5 (5.7)	& 26 (26)	& 11 (10)	& 7.0 (6.4)	\\ \midrule
  \multicolumn{2}{c}{NP limit}
		&  10.7		& 8.5		& 83.2		& 19.5		& 7.6		\\ 
  \multicolumn{2}{c}{Approximate $m_\phi$ limit (GeV)} 
		&  567 (574)	& 553 (556)	& 490 (489)	& 537 (532)	& 589 (579)	\\
  \bottomrule
 \end{tabular}
\caption{Acceptance times efficiency ($\mathcal{A}\varepsilon$) and total number of events ($N$) for three event samples without (with) pileup. The stop pair production sample is compared to the ATLAS result as a validation of our analysis. The 95\% CL limit on new physics (NP) contributions are given; these limits are quoted ATLAS results. Lastly we provide an approximate limit on $m_\phi$ based on our results.}
\label{tabSRtNbC}
\end{table*}

After preselection we demanded exactly two opposite sign leptons with the leading lepton having $p_T>25$~GeV, at least four jets with $p_T>80,60,40,25$~GeV, and at least one tagged b-jet.
We refer to Ref.~\cite{ATLAS-CONF-2013-037} for the definitions of the remaining kinematical variables and of the signal regions (SRs) SRtN2-3 and SRbC1-3, designed for $\tilde{t}_1\to t\tilde{\chi}_1^0$ and $\tilde{t}_1\to b\tilde{\chi}_1^\pm$ topologies respectively (see their Table~1). 
Variables $am_{T2}$ and $m_{T2}^\tau$ are variants of $m_{T2}$ designed to reject leptonic and semi-leptonic $t\bar{t}$ background respectively: $am_{T2}$ takes for its visible branches $b$ and $(bl)$, with a missing on-shell $W$ associated with the $b$ branch; $m_{T2}^\tau$ takes $l$ and a jet for its visible branches, assuming massless invisible states.
Both $am_{T2}$ and $m_{T2}^\tau$ require two jets in the event to be chosen as $b$-jets, regardless of whether they are $b$-tagged. ATLAS are able to choose those jets which have the highest $b$-tag weight. However, \textsc{Delphes} only outputs a boolean variable which identifies whether a jet is $b$-tagged or not. We must therefore find a way to choose two b-jets. We follow Ref.~\cite{Bai2012gs}. There are three cases:
\begin{itemize}
 \item 2 $b$-tags: Take both as $b$-jets.
 \item 1 $b$-tag: Assume that second $b$-jet is in the leading two non-$b$-tagged jets.
 \item 0 or $>2$ $b$-tags: Ignore $b$-tagging information and assume that $b$-jets are in leading three jets.
\end{itemize}
Then, to calculate $am_{T2}$, we take the $j_i(j_kl)$ permutation over the $b$-jet candidates which minimises $am_{T2}$. 
For $m_{T2}^\tau$ we assume that the $\tau$-jet is in the leading three jets. We find the $j_il$ combination over the candidate jets which minimises $m_{T2}^\tau$. These methods are in the spirit of $m_{T2}$ as the minimum mother particle mass consistent with the decay hypothesis and observed kinematics.
Since the minimum plausible $m_{T2}$ value is selected, the results after cuts are also conservative.
We compared our obtained $am_{T2}$ and $m_{T2}^\tau$ distributions for the $t\bar{t}$ sample at the preselection stage to Figure~3 in the ATLAS analysis \cite{ATLAS-CONF-2013-037} and found good agreement, particularly at large values where cuts are made.

The $N^{\text{iso-trk}}$ cut applied to the SRbC1-3 SRs cannot be replicated after our reconstruction has been performed. Cut-flows published in auxiliary Figures 112--117 of Ref.~\cite{Aad2014kra} suggest that after all other cuts, the $N^{\text{iso-trk}}$ requirement reduces the signal by 15--25\%, consistent between the single-muon and single-electron channel. We therefore conservatively post-scale our results in the SRbC1-3 SRs by a factor 0.75 to take this into account.

The results of our analysis are shown in Table~\ref{tabSRtNbC}. The acceptance times efficiency ($\mathcal{A}\varepsilon$) for our stop pair validation sample agree well with ATLAS results in each of the signal regions; our predicted event rates are likely an underestimate for the SRtN2-3 SRs. We are confident that the discrepancies can be assigned to some combination of: different event generators, the third-party detector simulation, our b-tagging efficiency approximation, the necessary amendments to $am_{T2}$ and $m_{T2}^\tau$ calculation methods, and our inability to recreate the pileup subtraction procedure. The predicted number of events in the 20.7~fb$^{-1}$ of data for each of the signal samples are also given in Table~\ref{tabSRtNbC}.

Since the branchings of $\phi$ and the distribution shapes do not change significantly from masses 500~GeV to 600~GeV, and since $\log[\sigma(pp\to\phi\phi)]$ varies approximately linearly with mass $m_\phi$, an approximate limit on $m_\phi$ can be determined by taking the published ATLAS new physics (NP) limits and assuming that, in each SR, the log of the number of accepted events scales linearly with $m_\phi$. These results are also shown in Table~\ref{tabSRtNbC}. Since this is only a recast of the ATLAS results, these limits are not to be taken too seriously; they serve only as an indication of the present experimental reach.

We note that these limits are found using the sum of single electron and muon channels. In our model $\approx 75\%$ of accepted events are single muons, whereas an approximately even share is expected for the background (and stops). We would likely obtain stronger limits if ATLAS published a NP limit on each lepton channel separately.

\subsubsection{$l^+l'^-\slashed{E}_T + jets$}

The dilepton final state is produced primarily through the mixed decay $\phi\phi \to b\nu tL \to l^+l'^- bb \slashed{E}_T$. 
There is also a non-negligible contribution from $\phi\phi \to tL tL'$, where the top pair and possible $\tau$ lepton(s) decay such that only two leptons are detected. 
This final state is the same as for leptonically decaying top pairs or for stop pairs decaying via the same chains considered in the previous subsection. ATLAS has performed a search for stop pairs in the dilepton final state, with no significant excess observed \cite{Aad2014qaa}. In this section we will recast the analysis in order to place a constraint on $m_\phi$.

\begin{table}\centering \footnotesize
 \begin{tabular}{ccc} 
  \toprule
  L110 & L100 & C1  \\
  \midrule
  \multicolumn{2}{c}{$m(l^+l^-)^{<71}_{>111}$~GeV} 	& opposite flavour	\\
  \multicolumn{2}{c}{$\Delta\phi_b<1.5$}  		& $m_{eff}>300$~GeV 	\\ 
  \multicolumn{2}{c}{$\Delta\phi_j>1.0$}  		& $\slashed{E}_T>50$~GeV  \\ 
\midrule
   - & $N(j)\ge 2$  					& $N(j)\ge 2$ \\
   - & $p_T^j[1]>100$~GeV  				& $p_T^j[1] > 50$~GeV	\\
     & $p_T^j[2]>50$~GeV   & \\ 
\midrule
  $m_{T2}>110$~GeV & $m_{T2}>100$~GeV 			& $m_{T2}>150$~GeV \\
  \bottomrule
 \end{tabular}
\caption{Signal region selections after preselection requirements.}
\label{tabSRs}
\end{table}

After preselection we demanded exactly two opposite sign leptons with the leading lepton having $p_T>25$~GeV. Any lepton pairs with invariant mass less than 20~GeV were rejected.
We then defined three SRs in Table~\ref{tabSRs}: L110, L100, and C1. We use the notation $p_T[1]$ ($p_T[2]$) to stand for the leading (subleading) $p_T$ object. 
We refer to Ref.~\cite{Aad2014qaa} for definitions of any unfamiliar variables. The most important is $m_{T2}$, which takes leptons for the visible branches and assumes massless missing particles. It is constructed to have a parton-level kinematic upper limit at $m_W$ for the dominant $t\bar{t}$ background.

\begin{figure}[t]\centering
 \includegraphics[width=\textwidth]{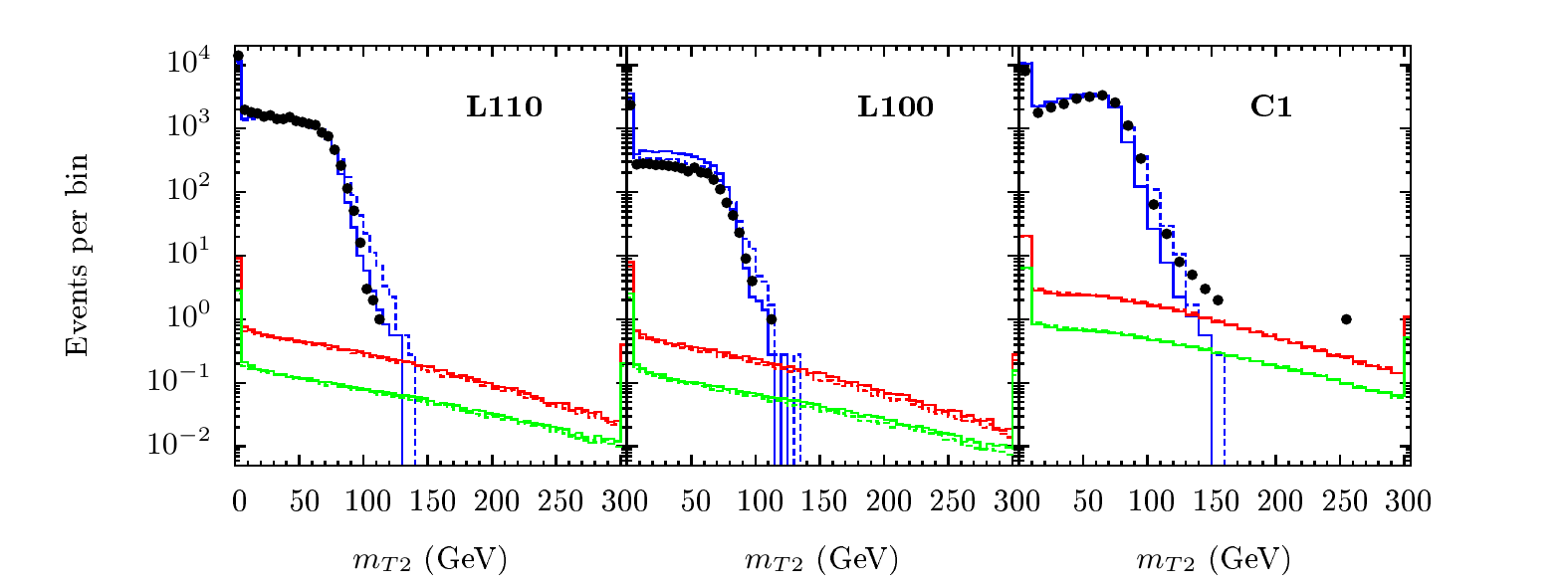}
	 \caption{Distribution of $m_{T2}$ opposite flavour events for the three SRs in $t\bar{t}$ and $m_\phi=500,600$~GeV event samples descending, simulated without (solid) and with (dashed) pileup. The ATLAS data, dominated by $t\bar{t}$ background for $m_{T2}\lesssim 100$~GeV, is overlaid as points. The apparent ``excess'' of $t\bar{t}$ events above $m_{T2}>100$~GeV is only because we have not simulated subleading backgrounds (only $t\bar{t}$ is necessary for validation of our analysis). These can be compared with Figures 9, 10, and 3 respectively of Ref.~\cite{Aad2014qaa}.}
 \label{fig_mT2OF}
\end{figure}

Plots of the number of events expected in 20.3~fb$^{-1}$ of integrated luminosity for each SR are shown against $m_{T2}$ for opposite flavour events in Fig.~\ref{fig_mT2OF}. These are to be compared with Figures 3, 9, and 10 of the ATLAS analysis \cite{Aad2014qaa}. One can see that our analysis does a good job of reproducing the background distribution in the region $m_{T2}\lesssim 100$~GeV where $t\bar{t}$ dominates. We are confident that the discrepancies can be assigned to some combination of: an overall normalisation factor, the LO $t\bar{t}$ event generator, the third-party detector simulation, and our inability to recreate the pileup subtraction procedure. The number of events in the SRs are broken up by lepton flavour in Table~\ref{tabSRlims}.

\begin{table*}\centering \footnotesize
 \begin{tabular}{ccccccc} 
  \toprule
		& \multicolumn{2}{c}{L110}	& \multicolumn{2}{c}{L100} 	& \multicolumn{2}{c}{C1}	\\ 
\midrule
  $m_\phi$~(GeV)& 500 & 600 			& 500 & 600 			& 500 & 600 			\\ 
\midrule
  $e^+e^-$	& 0.93 (0.86) & 0.34 (0.32) 	& 0.79 (0.68) & 0.30 (0.27) 	& - & -				\\
  $\mu^+\mu^-$	& 3.0 (2.8) & 1.0 (0.93)	& 2.7 (2.3) & 0.92 (0.81) 	& - & -				\\
  $\mu^\pm e^\mp$& 4.6 (4.2) & 1.5 (1.4)	& 3.9 (3.4) & 1.4 (1.2)		& 7.5 (7.6) & 2.8 (2.9)  \\ 
  $\sum_l l^+l'^-$& 8.5 (7.8) & 2.9 (2.7)	& 7.4 (6.3) & 2.6 (2.3)		& - & -  \\ \midrule
  NP limit	& \multicolumn{2}{c}{9.0}	& \multicolumn{2}{c}{5.6} 	& \multicolumn{2}{c}{2.3 $^*$} 	\\ 
  Approx. $m_\phi$ limit (GeV) & \multicolumn{2}{c}{ 495 (487) } & \multicolumn{2}{c}{ 527 (512) } & \multicolumn{2}{c}{ 621 (622) $^*$ } \\
\bottomrule
 \end{tabular}
\caption{Number of events in each SR without (with) pileup. The 95\% CL limit on new physics contributions are also given; these limits are quoted ATLAS results for L110 and L100, and inferred from a plot for C1 (which is why we mark it with a $^*$). }
\label{tabSRlims}
\end{table*}

The limits on the number of NP events summed over the lepton channels in SRs L110 and L100 are provided by ATLAS and reproduced in our Table~\ref{tabSRlims}. The limit from the C1 SR was not published, since this SR is subsequently filtered through a multivariate analysis. However, one can read off Figure~3 in Ref.~\cite{Aad2014qaa} that three events were observed with $3.6^{+6.7}_{-?}$ expected before the multivariate analysis. It is therefore reasonable to model the probability density function for the expected number of events as a gamma distribution with shape parameter 1.3 and mean 3.6.\footnote{The gamma distribution is the standard conjugate prior for rate parameters. A shape parameter of 1.3 ensures that $\int_{3.6}^{3.6+6.7}dx\,f(x;k=1.3,\mu=3.6)=34.1\%$, corresponding to one half of the 68.2\% confidence interval.} We performed toy Monte Carlo pseudoexperiments for different signal+background hypotheses ($H_{s+b}$) under this assumption, measuring
\begin{align}
 CL_s=\frac{Pr(n\le n_{obs} | H_{s+b} )}{Pr(n\le n_{obs} | H_b)}
\end{align}
each time. We found $CL_s=0.05$ for an expected new physics contribution of 2.3 events, corresponding to the observed $95\%$ CL limit on the number of NP events determined using the $CL_s$ method \cite{Read2002hq}, the same as that used in the ATLAS analysis.

An approximate limit on $m_\phi$ can be derived in the same way described in the previous subsection, and the results are shown in Table~\ref{tabSRlims}. Again, since this is only a recast of the ATLAS analysis, these limits are not to be taken too seriously; they serve only as an indication of the present experimental reach.

The best limit is obtained from the C1 SR. There are three principal reasons for this. \textit{(1)} The L110 and L100 limits are quoted on the sum over all flavour channels. In our model we expect greater than half of the events to be in the opposite-flavour channel. Simply requiring opposite flavour leptons reduces the background significantly (compare Figures 2 and 3 of Ref.~\cite{Aad2014qaa}), so that one can afford to make softer cuts that keep more signal. \textit{(2)} The L110 and L100 cuts on $\Delta \phi_b$ and $\Delta \phi_j$ are designed to reject background events with high $m_{T2}$ arising from events with large $\slashed{E}_T$ from mismeasured jets. These cuts keep about $50\%$ of the stop pair signals considered by ATLAS (see auxiliary Figures 24 and 25 of Ref.~\cite{Aad2014qaa}). We found that only $\approx 35\%$ of events were kept for our model due to different kinematics. \textit{(3)} The signal-to-background ratio and the limit is significantly improved if the cut on $m_{T2}$ is slightly 
increased.

\subsubsection{Summary}

It is clear from these analyses that the existing constraints on the leptoquark from sbottom and stop searches are comparable, $m_\phi\gtrsim 600$~GeV.
Inferred limits could be even stronger if the collaborations provided limits before combining lepton flavour channels.
But this conclusion can be turned around: if the collaborations \textit{were} to see a significant excess in any of the discussed final states, 
this model predicts that it should show up in all of them at around the same time, with a well-predicted, non-universal flavour signature
and distinctive kinematics.
Simple SUSY models might find this scenario difficult to accommodate.

\section{Conclusion}
\label{sec:con}
In this work we have written down the minimal UV completions for all of the D7 $\Delta L=2$ operators 
which could be responsible for radiatively generating a Majorana neutrino mass.
We then discussed the generic collider searches for the newly introduced exotic particles, 
including vector-like quarks, vector-like leptons, scalar leptoquarks, a charged scalar, a scalar doublet and a scalar quadruplet.
The properties of these particles are generally constrained by low-energy neutrino oscillation data.
The hope is that this will advance a systematic approach to searches for the origin of neutrino mass at the LHC.

A detailed study of the collider bounds has been presented for 
$\mathcal{O}_3=LLQ\bar{d}H$ and $\mathcal{O}_8=L\bar{d}\bar{e}^\dagger\bar{u}^\dagger H$ completions 
where a leptoquark $\phi\sim (\bar{3},1,\frac{1}{3})$ and a vector-like quark $\chi\sim (3,2,-\frac{5}{6})$ are introduced.
In the detailed study, we constrained the vector-like quark mass $m_\chi \gtrsim 620$~GeV using a dedicated LHC search.
For the leptoquark $\phi$ we recast LHC sbottom/stop searches and explored in the parameter space allowed by the constraints from flavour physics.
We found two distinct areas of parameter space, one where Br$(\phi\to b\nu)\approx 100\%$ and the other where Br$(\phi\to b\nu)<100\%$.
In the first case $m_\phi\gtrsim 520$--$600$~GeV, and in the second case we found $m_\phi \gtrsim 600$~GeV using three different final states.

Through this detailed analysis we have shown the powerful discovery and/or exclusion potential of the LHC for the radiative neutrino mass models
based on $\Delta L=2$ operators.
We have also made advances in a systematic approach to these searches.
  
 \section*{Acknowledgements}
We thank Aldo Saavedra for useful discussions. This work was supported in part by the Australian Research Council.


\appendix

\section{Details of Models}
\label{app:uvcompletion}
The details of the minimal UV-completion of the D7 operators of Eq.~\ref{eq:d7operators} are described here.
New fermions are denoted by $\chi$. As it turns out that all of them have a non-vanishing hypercharge, they have to be accompanied by a Dirac partner $\bar\chi$, in order to be able to write down a mass term. New scalars are labelled $\phi$ and the charge conjugate is denoted  $\tilde \phi$ analogously to the SM Higgs $H$. We will suppress the SU(2)$_L$ contractions whenever they are unique, e.g. $LL \equiv L^\alpha L^\beta \epsilon_{\alpha\beta}$. We use roman indices $i,j,\dots$ for flavour, greek indices $\alpha,\beta,\dots$ for SU(2)$_L$ and roman indices $a,b,\dots$ for SU(3)$_c$.
The collider searches for relevant exotic particles are discussed in Sec.~\ref{sec:lhcsearches}.
\subsection{Operator $\mathcal{O}_2=LLL\bar e H$}
\begin{figure}[b]
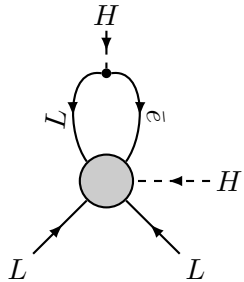
\centering
\FMDG{O2nuMass}
\caption{Neutrino mass from $\mathcal{O}_2$.}
\label{fig:O2nuMass}
\end{figure}

The structure of the operator is as follows,
\begin{equation}
\mathcal{O}_2 = L^\alpha L^\beta L^\gamma \bar{e} H^\delta \epsilon_{\alpha\beta}\epsilon_{\gamma\delta} ,
\end{equation}
and the corresponding neutrino mass diagram is shown in \Figref{fig:O2nuMass}.
The flavour structure of the operator is
\begin{equation}
\kappa^{\mathcal{O}_2}_{ijkl} L^\alpha_i L^\beta_j L^\gamma_k \bar e_l H^\delta\epsilon_{\alpha\beta}\epsilon_{\gamma\delta} ,
\end{equation}
where the first two indices are anti-symmetric. 
At low-energy, the neutrino mass is given by 
\begin{equation}
\left(m_\nu\right)_{ij} \simeq \kappa^{\mathcal{O}_2}_{[ik] j k} m_{\ell_k} \, I 
\end{equation}
with the loop integral $I$. The proportionality on the charged lepton mass $m_{\ell_k}$ introduces a hierarchy which might require more generations of new particles unless it can be compensated by other Yukawa couplings. 
The minimal UV completions are shown in \Tabref{tab:O2}.
\begin{table}[bt]
\begin{center}
\begin{tabular}{cc}
\toprule
& $\mathcal{O}_2^1$  \\
& $L (LL) (\bar{e} H)$ \\
\midrule
$\phi$ &   $(1, 1, 1)$ \\[1ex]
$\chi$ &   $(1, 2, -\frac{3}{2})$ \\
\bottomrule
\end{tabular}
\hspace{5ex}
\begin{tabular}{cc}
\toprule
& $\mathcal{O}_2^2$\\
 & $ H (LL) (L \bar{e})$\\
\midrule
     $\phi$ & $(1, 1, 1)$ \\[1ex]
  $\eta$ & $(1, 2, \frac{1}{2})$ \\
\bottomrule
\end{tabular}
\end{center}
\caption{Minimal UV completions of operator $\mathcal{O}_2$.} 
\label{tab:O2}
\end{table}

\tocless\subsubsection{$\mathcal{O}_2^1$ Model}
The model contains a complex scalar $\phi$ and a Dirac fermion $\chi+\bar{\chi}$ with quantum numbers
\begin{equation}
\phi\sim(1,1,1),\qquad \chi\sim\left(1,2,-\frac32\right)\;.
\end{equation}
The Lagrangian $\Delta\mathcal{L}= \mathcal{L}_Y -\mathcal{V}$ is given by 
\begin{align}
- \mathcal{L}_Y &= m_\chi \bar\chi\chi+ Y^{LL\phi}_{ij} L_iL_j\phi + Y^{L\chi\phi}_{i} L_i \bar{\chi} \phi^\dagger + Y^{\bar e\chi H}_{i} \bar{e}_i \chi H +\hc \\
\mathcal{V}&=\mu^2_{\phi} \phi^\dagger\phi
+ \lambda_\phi (\phi^\dagger\phi)^2
+ \lambda_{H\phi} H^\dagger H \phi^\dagger \phi \;.
\end{align}
Note that the singly charged component of the new Dirac fermion $\chi+\bar{\chi}$ mixes with the SM leptons, which introduces non-unitarity to the ordinary $3\times 3$ PMNS mixing matrix.


\tocless\subsubsection{$\mathcal{O}_2^2$ Model}
This model is called Zee model~\cite{Zee:1980ai}, which introduces two scalar fields, and its particle content is given by
\begin{equation}
\phi\sim(1,1,1),\qquad \eta\sim\left(1,2,\frac12\right)\;.
\end{equation}
The Lagrangian $\Delta\mathcal{L}= \mathcal{L}_Y -\mathcal{V}$ is  given by 
\begin{align}
 -\mathcal{L}_Y &= Y^{LL\phi}_{ij} L_iL_j \phi + Y^{L\bar e\eta}_{ij} L_i\bar e_j \tilde\eta + Y^{Q\bar d\eta}_{ij} Q_i\bar d_j \tilde\eta + Y^{Q\bar u\eta}_{ij} Q_i\bar u_j \eta+\hc \\
 \mathcal{V}&= V_\text{2HDM}(H,\eta)
+ \mu^2_{\phi} \phi^\dagger\phi
+ \lambda_{\phi} (\phi^\dagger\phi)^2
+\lambda_{H\phi} H^\dagger H\phi^\dagger\phi 
+\lambda_{\eta\phi} \eta^\dagger \eta\phi^\dagger\phi \\\nonumber
& + (\kappa_{H\eta\phi} H\eta\phi^\dagger +\lambda_{H\eta\phi} H^\dagger \eta\phi^\dagger\phi +\hc) \;,
\end{align}
where $V_\text{2HDM}(H,\eta)$ denotes the general two Higgs doublet model potential.

\subsection{Operator $\mathcal{O}_3=LLQ\bar d H$}
There are two possible SU(2)$_L$ structures of the operator $\mathcal{O}_3$,
\begin{align}
\mathcal{O}_{3a} &= L^\alpha L^\beta Q^\gamma \bar{d} H^\delta \epsilon_{\alpha\beta}\epsilon_{\gamma\delta}, & 
\mathcal{O}_{3b} &= L^\alpha L^\beta Q^\gamma \bar{d} H^\delta \epsilon_{\alpha\gamma}\epsilon_{\beta\delta},
\end{align}
and the corresponding neutrino mass diagrams are shown in \Figref{fig:O3nuMass}. 
The flavour structure of the operator $\mathcal{O}_{3a}$ is
\begin{equation}
\kappa^{\mathcal{O}_{3a}}_{ijkl} L_i^\alpha L_j^\beta Q_k^\gamma \bar d_l H^\delta\epsilon_{\alpha\beta}\epsilon_{\gamma\delta},
\end{equation}
where the first two indices are anti-symmetric. 
For operator $\mathcal{O}_{3a}$ we obtain neutrino masses
\begin{equation}
(m_\nu)_{ij} \simeq  g_2^2 \kappa^{\mathcal{O}_{3a}}_{[ij] kl} (m_d)_{kl} \,I 
\end{equation}
with the loop integral $I$. The flavour structure of the operator $\mathcal{O}_{3b}$ is
\begin{equation}
\kappa^{\mathcal{O}_{3b}}_{ijkl} L^\alpha_i L^\beta_j Q^\gamma_k \bar d_l H^\delta \epsilon_{\alpha\gamma}\epsilon_{\beta\delta}
\end{equation} 
and the neutrino mass matrix is
\begin{equation}
(m_\nu)_{ij} \simeq \kappa^{\mathcal{O}_{3b}}_{ijkl} (m_d)_{kl}  \, I 
\end{equation}
with the loop integral $I$. The proportionality on the down-type quark mass matrix $m_d$ introduces a hierarchy which might require more generations of new particles.
\begin{figure}[bt]\centering
\begin{subfigure}{10cm}\centering
\FMDG{O3anuMass}
\FMDG{O3anuMass2}
\caption{Neutrino mass from $\mathcal{O}_{3a}$.}
\label{fig:O3anuMass}
\end{subfigure}
\begin{subfigure}{5cm}\centering
\FMDG{O3bnuMass}
\caption{Neutrino mass from $\mathcal{O}_{3b}$.}
\label{fig:O3bnuMass}
\end{subfigure}
\caption{Neutrino mass from $\mathcal{O}_3$.}
\label{fig:O3nuMass}
\end{figure}
The relevant minimal UV completions are collected in \Tabref{tab:O3}. There can not be any colour adjoint representations, because this requires 4 quarks in the operator.
\begin{table}[bt]
\begin{center}
   \begin{tabular}{ccccccc}
   \toprule
& $\mathcal{O}_{3}^{1}$ &  $\mathcal{O}_{3}^{2}$ & $\mathcal{O}_{3}^{3}$ & $\mathcal{O}_{3}^{4}$ & $\mathcal{O}_{3}^{5}$ & $\mathcal{O}_{3}^{6}$\\
     & $Q(LL) (\bar{d}H)$ & $\bar{d} (LL) (Q H)$ & $L (L\bar{d}) (QH)$ & $L(LQ) (\bar{d} H)$  & $L(LQ) (\bar{d} H)$ & $L (L\bar{d}) (QH)$\\
      \midrule
      $\phi$    & $(1, 1, 1)$ &   $(1, 1, 1)$& $(3,2,\frac{1}{6})$ & $(3, 1, -\frac{1}{3})$  & $(3, 3, -\frac{1}{3})$ & $(3,2,\frac{1}{6})$  \\[1ex]
      $\chi$  &  $(3, 2, -\frac{5}{6})$& $(3, 1, \frac{2}{3})$ & $(3, 1, \frac{2}{3})$ & $(3, 2, -\frac{5}{6})$  & $(3, 2, -\frac{5}{6})$  & $(3, 3, \frac{2}{3})$ \\
      \midrule
&$\mathcal{O}_{3a}$&$\mathcal{O}_{3a}$&$\mathcal{O}_{3a}$&$\mathcal{O}_{3b}$&$\mathcal{O}_{3a},\mathcal{O}_{3b}$&$\mathcal{O}_{3a},\mathcal{O}_{3b}$\\
\bottomrule
   \end{tabular}

\vspace{2ex}
   \begin{tabular}{cccc}
   \toprule
 & $\mathcal{O}_{3}^{7}$ &  $\mathcal{O}_{3}^{8}$ &  $\mathcal{O}_{3}^{9}$\\
   & $H (LL) (Q \bar{d})$& $H (LQ) (L\bar{d})$& $H (LQ) (L\bar{d})$\\
   \midrule
    $\phi$ & $(1,1,1)$&  $(3, 1, -\frac{1}{3})$&  $(3, 3, -\frac{1}{3})$\\[1ex]
    $\eta$  & $(1, 2, \frac{1}{2}) $   &  $(3,2,\frac{1}{6})$  &  $(3,2,\frac{1}{6})$\\ 
    \midrule
&$\mathcal{O}_{3a}$&$\mathcal{O}_{3b}$&$\mathcal{O}_{3a},\mathcal{O}_{3b}$\\
\bottomrule
   \end{tabular}

\end{center}
\caption{Minimal UV completions of operators $\mathcal{O}_{3a}$ and $\mathcal{O}_{3b}$.}
\label{tab:O3}
\end{table}

\tocless\subsubsection{$\mathcal{O}_{3}^{1}$ Model}
The model contains a complex scalar $\phi$ and a Dirac fermion $\chi+\bar{\chi}$ with quantum numbers
\begin{equation}
\phi\sim(1,1,1),\qquad \chi\sim\left(3,2,-\frac56\right)\;.
\end{equation}
The additional interaction Lagrangian $\Delta\mathcal{L}=\mathcal{L}_Y-\mathcal{V}$ is given by
\begin{align}
-\mathcal{L}_Y&=m_\chi \bar\chi\chi+ Y^{LL\phi}_{ij} L_i L_j \phi 
+Y^{Q\bar \chi\phi}_{i} Q_i\bar\chi \phi^\dagger 
+Y^{\bar d\chi H}_{i} \bar d_i \chi H
+ \hc\\
\mathcal{V}&=\mu^2_\phi \phi^\dagger\phi
+ \lambda_\phi (\phi^\dagger\phi)^2
+\lambda_{H\phi} H^\dagger H \phi^\dagger \phi\;.
\end{align}
This model leads to the effective operator $\mathcal{O}_{3a}$.

\tocless\subsubsection{$\mathcal{O}_{3}^{2}$ Model}
The model contains a complex scalar $\phi$ and a Dirac fermion $\chi+\bar{\chi}$ with quantum numbers
\begin{equation}
\phi\sim(1,1,1),\qquad \chi\sim\left(3,1,\frac23\right)\;.
\end{equation}
The additional interaction Lagrangian $\Delta\mathcal{L}=\mathcal{L}_Y-\mathcal{V}$ is given by
\begin{align}
-\mathcal{L}_Y&=m_\chi \bar\chi\chi+m_{\bar u\chi,i} \bar u_i \chi + Y^{LL\phi}_{ij} L_i L_j \phi
+Y^{Q\chi H}_{ij} Q_i\bar\chi H
+Y^{\bar d\chi \phi}_{i} \bar d_i \chi \phi^\dagger
+ \hc\\
\mathcal{V}&=\mu^2_\phi\phi^\dagger\phi
+ \lambda_\phi (\phi^\dagger\phi)^2
+\lambda_{H\phi} H^\dagger H \phi^\dagger \phi\;.
\end{align}
This model leads to the effective operator $\mathcal{O}_{3a}$.

\tocless\subsubsection{$\mathcal{O}_{3}^{3}$ Model}
The model contains a complex scalar $\phi$ and a Dirac fermion $\chi+\bar{\chi}$ with quantum numbers
\begin{equation}
\phi\sim\left(3,2,\frac16\right),\qquad \chi\sim\left(3,1,\frac23\right)\;.
\end{equation}
This model has been studied in Ref.~\cite{Babu:2011vb}.
The additional interaction Lagrangian $\Delta\mathcal{L}=\mathcal{L}_Y-\mathcal{V}$ is given by
\begin{align}
-\mathcal{L}_Y&=m_\chi \bar\chi\chi+m_{\bar u\chi,i} \bar u_i \chi
+ Y^{L\bar d\phi}_{ij} L_i \bar d_j \phi
+Y^{Q\chi H}_{i} Q_i\bar\chi H
+Y^{L\chi \phi}_{i} L_i \chi \phi^\dagger
+ \hc\\
\mathcal{V}&=\mu^2_\phi\phi^\dagger\phi
+ \lambda_\phi (\phi^\dagger\phi)^2
+\lambda_{H\phi} H^\dagger H \phi^\dagger \phi\;.
\end{align}
Couplings such as $H^\dagger \phi^3 $
vanish due to the antisymmetric nature of the colour contraction
unless more than one copy of the scalar fields are introduced. 
We shall not comment on this type of couplings from now on to avoid      
unnecessarily repeated discussion.  
This model leads to the effective operator $\mathcal{O}_{3a}$.

\tocless\subsubsection{$\mathcal{O}_{3}^{4}$ Model}
The model contains a complex scalar $\phi$ and a Dirac fermion $\chi+\bar{\chi}$ with quantum numbers
\begin{equation}
\phi\sim\left(3,1,-\frac13\right),\qquad \chi\sim\left(3,2,-\frac56\right)\;.
\end{equation}
The additional interaction Lagrangian $\Delta\mathcal{L}=\mathcal{L}_Y-\mathcal{V}$ is given by
\begin{align}
-\mathcal{L}_Y&=m_\chi \bar\chi\chi
+Y^{LQ\phi}_{ij} L_i Q_j \phi^\dagger
+Y^{L\chi\phi}_{i} L_i\bar\chi \phi
+ Y^{QQ\phi}_{ij} Q_iQ_j\phi \\\nonumber
&+ Y^{\bar d\chi H}_{i} \bar d_i \chi H 
+ Y^{\bar d\bar u\phi}_{ij} \bar d_i \bar u_j \phi^\dagger 
+Y^{\bar e\bar u \phi}_{ij} \bar e_i \bar u_j \phi
+ \hc\\
\mathcal{V}&=
\mu^2_\phi \phi^\dagger\phi
+ \lambda_\phi (\phi^\dagger\phi)^2
+\lambda_{H\phi} H^\dagger H \phi^\dagger \phi\;.
\end{align}
This model leads to the effective operator $\mathcal{O}_{3b}$.

\tocless\subsubsection{$\mathcal{O}_{3}^{5}$ Model}
The model contains a complex scalar $\phi$ and a Dirac fermion $\chi+\bar{\chi}$ with quantum numbers
\begin{equation}
\phi\sim\left(3,3,-\frac13\right),\qquad \chi\sim\left(3,2,-\frac56\right)\;.
\end{equation}
The additional interaction Lagrangian $\Delta\mathcal{L}=\mathcal{L}_Y-\mathcal{V}$ is given by
\begin{align}
-\mathcal{L}_Y&=
m_\chi \bar\chi\chi
+Y^{LQ\phi}_{ij} L_i^\alpha Q_j^\gamma \left( \epsilon_{\alpha\beta} \phi^\dagger_{\beta\gamma}   + \epsilon_{\gamma\beta}\phi^\dagger_{\beta\alpha}  \right)
+Y^{L\chi\phi}_{i} L_i^\alpha \bar\chi^\gamma \left(  \epsilon_{\alpha\beta}\phi_{\beta\gamma}+\epsilon_{\gamma\beta}\phi_{\beta\alpha}\right)  \nonumber\\
& + Y^{QQ\phi}_{ij} Q_i^\alpha  Q_j^\gamma \epsilon_{\alpha\beta}\phi_{\beta\gamma}
+ Y^{\bar d\chi H}_{i} \bar d_i \chi H 
+ \hc\\
\mathcal{V}&=\mu^2_\phi \tr\phi^\dagger\phi
+ \lambda_{\phi,1} (\tr \phi^\dagger\phi)^2
 + \lambda_{\phi,2} \tr([\phi^\dagger,\phi]^2)
+\lambda_{H\phi,1} H^\dagger H \tr (\phi^\dagger \phi )
+\lambda_{H\phi,2} H^\dagger [\phi^\dagger, \phi] H
\;.
\end{align}
Both operators $\mathcal{O}_{3a}$ and $\mathcal{O}_{3b}$ are generated.

\tocless\subsubsection{$\mathcal{O}_{3}^{6}$ Model}
The model contains a complex scalar $\phi$ and a Dirac fermion $\chi+\bar{\chi}$ with quantum numbers
\begin{equation}
\phi\sim\left(3,2,\frac16\right),\qquad \chi\sim\left(3,3,\frac23\right)\;.
\end{equation}
The additional interaction Lagrangian $\Delta\mathcal{L}=\mathcal{L}_Y-\mathcal{V}$ is given by
\begin{align}
-\mathcal{L}_Y&=m_\chi\tr\left( \bar\chi\chi\right)
+Y^{L\bar d\phi}_{ij} L_i \bar d_j \phi
+Y^{L\chi\phi}_{i} L_i^\alpha \left(\epsilon_{\alpha\beta}\chi_{\beta\gamma}+\epsilon_{\gamma\beta}\chi_{\beta\alpha}\right)\phi^\dagger_\gamma\\\nonumber
&+Y^{Q\chi H}_{i} Q_i^\alpha\left(\epsilon_{\alpha\beta} \bar\chi_{\beta\gamma}+\epsilon_{\gamma\beta} \bar\chi_{\beta\alpha} \right)H^\gamma
+ \hc\\
\mathcal{V}&=\mu^2_\phi \phi^\dagger\phi
 + \lambda_{\phi} (\phi^\dagger\phi)^2
+\lambda_{H\phi} H^\dagger H \phi^\dagger \phi\;.
\end{align}
Both operators $\mathcal{O}_{3a}$ and $\mathcal{O}_{3b}$ are generated.

\tocless\subsubsection{$\mathcal{O}_{3}^{7}$ Model}
This model is exactly the same as $\mathcal{O}_2^2$. It also leads to the effective operator $\mathcal{O}_{3a}$.

\tocless\subsubsection{$\mathcal{O}_{3}^{8}$ Model}
The model contains two additional complex scalars $\phi$ and $\eta$,
\begin{equation}
\phi\sim\left(3, 1,-\frac13\right),\qquad \eta\sim\left(3,2,\frac16\right)\;.
\end{equation}
This model was given as an example in Ref.~\cite{Babu:2001ex} and studied in Ref.~\cite{Babu:2010vp}.
The additional interaction Lagrangian $\Delta\mathcal{L}=\mathcal{L}_Y-\mathcal{V}$ is given by
\begin{align}
-\mathcal{L}_Y&=Y^{L\bar d\eta}_{ij} L_i \bar d_j \eta
+Y^{LQ\phi}_{ij} L_i Q_j\phi^\dagger
+Y^{QQ\phi}_{ij} Q_i Q_j \phi
+Y^{\bar d \bar u\phi}_{ij} \bar d_i \bar u_j \phi^\dagger
+Y^{\bar e \bar u\phi}_{ij} \bar e_i \bar u_j \phi
+ \hc\\
\mathcal{V}&=\mu_\phi^2 \phi^\dagger\phi + \mu_\eta^2 \eta^\dagger\eta +\lambda_\phi (\phi^\dagger\phi)^2+\lambda_\eta (\eta^\dagger\eta)^2 +\lambda_{H\phi} H^\dagger H \phi^\dagger\phi +\lambda_{H\eta} H^\dagger H \eta^\dagger\eta \\\nonumber
&+\lambda_{\phi\eta} \phi^\dagger\phi \eta^\dagger\eta + (\kappa\, \eta^\dagger H \phi +\hc)\;.
\end{align}
This model leads to the effective operator $\mathcal{O}_{3b}$.

\tocless\subsubsection{$\mathcal{O}_{3}^{9}$ Model}
The model contains two additional complex scalars $\phi$ and $\eta$
\begin{equation}
\phi\sim\left(3, 3,-\frac13\right),\qquad \eta\sim\left(3,2,\frac16\right)\;.
\end{equation}
This model was given as an example in Ref.~\cite{Babu:2001ex}.
The additional interaction Lagrangian $\Delta\mathcal{L}=\mathcal{L}_Y-\mathcal{V}$ is given by
\begin{align}
-\mathcal{L}_Y&=Y^{L\bar d\eta}_{ij} L_i \bar d_j \eta
+Y^{LQ\phi}_{ij} L_i^\alpha Q_j^\gamma \left(\epsilon_{\alpha\beta}\phi^\dagger_{\beta\gamma}  +\epsilon_{\gamma\beta}\phi^\dagger_{\beta\alpha}\right)
+Y^{QQ\phi}_{ij} Q_i^\alpha Q_j^\gamma\epsilon_{\alpha\beta}\phi_{\beta\gamma}
+ \hc\\
\mathcal{V}&=\mu_\phi^2\,\tr \phi^\dagger \phi 
+ \mu_\eta^2\, \eta^\dagger\eta 
+ \lambda_{\phi,1} \left(\tr\phi^\dagger \phi\right)^2 
+ \lambda_{\phi,2} \tr([\phi^\dagger, \phi]^2) 
+ \lambda_{\eta} (\eta^\dagger \eta)^2
+ \lambda_{H\eta} H^\dagger H \eta^\dagger \eta\nonumber\\
& + \lambda_{H\phi,1} H^\dagger H \tr\phi^\dagger \phi 
+ \lambda_{H\phi,2} H^\dagger [\phi^\dagger, \phi ] H
+ \lambda_{\phi\eta,1} \eta^\dagger \eta \tr \phi^\dagger\phi
+ \lambda_{\phi\eta,2} \eta^\dagger [\phi^\dagger,\phi]\eta\\\nonumber
&+ \Big(\kappa\, \eta^\dagger_\alpha\left(\epsilon_{\alpha\beta} \phi_{\beta\gamma}+\epsilon_{\gamma\beta} \phi_{\beta\alpha}\right) H_\gamma 
+ \kappa^\prime \, \epsilon_{abc}\eta^a_\alpha\eta^b_\gamma \left(\epsilon_{\alpha\beta} \phi^c_{\beta\gamma}-\epsilon_{\gamma\beta} \phi^c_{\beta\alpha}\right) \\\nonumber
&+\tilde\kappa\, \epsilon_{abc} \eta^a_\alpha \epsilon_{\alpha\beta}[\phi^b,\phi^c]_{\beta\gamma} H_\gamma + \hc\Big)\;.
\end{align}
Both operators $\mathcal{O}_{3a}$ and $\mathcal{O}_{3b}$ are generated.

\subsection{Operator $\mathcal{O}_{4} =L L Q^\dagger \bar{u}^\dagger H$}
There are two possible SU(2)$_L$ structures of the operator are given by
\begin{align}
\mathcal{O}_{4a}&=L^\alpha  L^\beta   (Q^\dagger)^\gamma  \bar u^\dagger H^\delta \epsilon_{\alpha\gamma}\epsilon_{\beta\delta} ,&
\mathcal{O}_{4b}&=L^\alpha  L^\beta  (Q^\dagger)^\gamma  \bar u^\dagger H^\delta \epsilon_{\alpha\beta}\epsilon_{\gamma\delta}\;.
\end{align}
The corresponding neutrino mass diagram can be obtained from the ones in \Figref{fig:O3bnuMass} (\Figref{fig:O3anuMass}) for $\mathcal{O}_{4b}$ ($\mathcal{O}_{4a}$) by replacing $\bar d$ by $\bar u$ and reversing the arrows of the quark lines. 
The flavour structure of the operator $\mathcal{O}_{4a}$ is
\begin{equation}
\kappa^{\mathcal{O}_{4a}}_{ijkl} L_i^\alpha L_j^\beta (Q^\dagger)_k^\gamma \bar u^\dagger_l H^\delta \epsilon_{\alpha\gamma}\epsilon_{\beta\delta} ,
\end{equation} 
and thus the neutrino mass matrix is
\begin{equation}
(m_\nu)_{ij} \simeq \kappa^{\mathcal{O}_{4a}}_{ijkk} m_{u_k} \, I 
\end{equation}
with the loop integral $I$. The proportionality to the up-type quark mass $m_{u_k}$ introduces a hierarchy which might require more generations of new particles unless it can be compensated by other Yukawa couplings. 

The flavour structure of the operator $\mathcal{O}_{4b}$ is
\begin{equation}
\kappa^{\mathcal{O}_{4b}}_{ijkl} L_i^\alpha L_j^\beta (Q^\dagger)_k^\gamma \bar u^\dagger_l H^\delta\epsilon_{\alpha\beta}\epsilon_{\gamma\delta}, 
\end{equation}
where the first two indices are anti-symmetric. 
The neutrino mass matrix is given by
\begin{equation}
(m_\nu)_{ij} \sim  g_2^2 \kappa^{\mathcal{O}_{4b}}_{[ij] kk} \,m_{u_k} \,I 
\end{equation}
with the loop integral $I$. The proportionality on the up-type quark mass $m_{u_k}$ introduces a hierarchy which might require more generations of new particles unless it can be compensated by other Yukawa couplings. 
The UV completions are listed in Tab.~\ref{tab:O4}. There can not be any colour adjoint representations, because this requires 4 quarks in the operator.
\begin{table}[tb]
\centering
\begin{tabular}{ccc}
\toprule
 & $\mathcal{O}_{4}^{1}$ &  $\mathcal{O}_{4}^{2}$\\
&   $Q^\dagger (LL)(\bar{u}^\dagger H)$ & $\bar{u}^\dagger (LL)(Q^\dagger H)$\\
\midrule
$\phi$ &  $(1,1,1)$ & $(1,1,1)$\\
$\chi^\dagger$ &  $(3,2,\frac{7}{6})$ & $(3,1,-\frac{1}{3})$\\[1ex]
\midrule
 & $\mathcal{O}_{4b}$& $\mathcal{O}_{4b}$\\
\bottomrule
\end{tabular}
\hspace{5ex}
\begin{tabular}{cc}
\toprule
& $\mathcal{O}_{4}^{3}$ \\
& $H(LL)(Q^\dagger \bar{u}^\dagger) $ \\
\midrule
$\phi$ &  $(1,1,1)$\\[1ex]
$\eta$ &  $(1,2,\frac{1}{2})$\\ 
\midrule
& $\mathcal{O}_{4b}$\\
\bottomrule
\end{tabular}
\caption{Minimal UV completions of the operators $\mathcal{O}_{4a}$ and $\mathcal{O}_{4b}$. Only the operator $\mathcal{O}_{4b}$ is generated.}
\label{tab:O4}
\end{table}


\tocless\subsubsection{$\mathcal{O}_{4}^{1}$ Model}
The model contains a complex scalar $\phi$ and a Dirac fermion $\chi+\bar{\chi}$ with quantum numbers
\begin{equation}
\phi\sim(1,1,1),\qquad \chi\sim\left(3,2,\frac76\right)\;.
\end{equation}
The additional interaction Lagrangian $\Delta\mathcal{L}=\mathcal{L}_Y-\mathcal{V}$ is given by
\begin{align}
-\mathcal{L}_Y&=m_\chi \bar\chi\chi+Y^{LL\phi}_{ij} L_i L_j \phi 
+Y^{Q \bar \chi\phi}_{i} Q_i\bar\chi \phi
+Y^{\bar u\chi H}_{i} \bar u_i \chi \tilde H
+ \hc\\
\mathcal{V}&=\mu^2_\phi\phi^\dagger\phi
+ \lambda_\phi (\phi^\dagger\phi)^2
+\lambda_{H\phi} H^\dagger H \phi^\dagger \phi\;.
\end{align}
This model leads to operator $\mathcal{O}_{4b}$.

\tocless\subsubsection{$\mathcal{O}_{4}^{2}$ Model}
The model contains a complex scalar $\phi$ and a Dirac fermion $\chi+\bar{\chi}$ with quantum numbers
\begin{equation}
\phi\sim(1,1,1),\qquad \chi\sim\left(3,1,-\frac13\right)\;.
\end{equation}
The additional interaction Lagrangian $\Delta\mathcal{L}=\mathcal{L}_Y-\mathcal{V}$ is given by
\begin{align}
-\mathcal{L}_Y&=m_\chi \bar\chi\chi+m_{\bar d\chi,i} \bar d_i\chi+Y^{LL\phi}_{ij} L_i L_j \phi 
+Y^{Q\chi H}_{i} Q_i\bar\chi \tilde H
+Y^{\bar u\chi \phi}_{i} \bar u_i \chi \phi
+ \hc\\
\mathcal{V}&=\mu^2_{\phi}\phi^\dagger\phi
+ \lambda_\phi (\phi^\dagger\phi)^2
+\lambda_{H\phi} H^\dagger H \phi^\dagger \phi\;.
\end{align}
This model leads to operator $\mathcal{O}_{4b}$.

\tocless\subsubsection{$\mathcal{O}_{4}^{3}$ Model}
This model is the same model as $\mathcal{O}_{3}^{7}$ and $\mathcal{O}_{2}^{2}$. It also leads to operator $\mathcal{O}_{4b}$.

\subsection{Operator $\mathcal{O}_8=L \bar{d} \bar{e}^\dagger \bar{u}^\dagger H$}
\begin{figure}[bt]
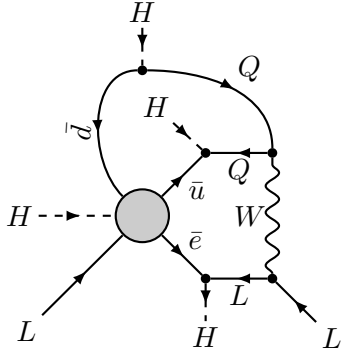
\centering
\FMDG{O8nuMass}
\caption{Neutrino mass from $\mathcal{O}_{8}$.}
\label{fig:O8nuMass}
\end{figure}

The SU(2)$_L$ structure of the operator is given by
\begin{equation}
L^\alpha\bar d\bar e^\dagger \bar u^\dagger H^\beta \epsilon_{\alpha\beta}
\end{equation}
and the corresponding neutrino mass diagram is shown in \Figref{fig:O8nuMass}. 
The flavour structure of the operator is
\begin{equation}
\kappa^{\mathcal{O}_{8}}_{ijkl} L_i^\alpha\bar d_j \bar e^\dagger_k \bar u^\dagger_l H^\beta \epsilon_{\alpha\beta}
\end{equation} 
and the neutrino mass matrix is
\begin{equation}
(m_\nu)_{ij} \simeq \kappa^{\mathcal{O}_{8}}_{ik jl} (m_d^\dagger m_u)_{kl} m_{\ell_j}  \, I 
\end{equation}
with the loop integral $I$. The proportionality to the product of the charged fermion mass matrices $(m_d^\dagger m_u)_{kl} m_{\ell_j}$ introduces a hierarchy, which might require more generations of new particles unless it can be compensated by other Yukawa couplings.
The UV completions are listed in Tab.~\ref{tab:O8}.
\begin{table}[tb]
\centering
\begin{tabular}{cccc}
\toprule
 &$\mathcal{O}_{8}^{1}$ &$\mathcal{O}_{8}^{2}$ &  $\mathcal{O}_{8}^{3}$ \\
 & $L(\bar e^\dagger \bar u^\dagger) (\bar d H)$  & $\bar{u}^\dagger (L\bar{d})(\bar{e}^\dagger H)$ 
& $\bar{e}^\dagger (L\bar{d})(\bar{u}^\dagger H)$  \\
\midrule
$\phi$          & $(3,1,-\frac13)$  &  $(3,2,\frac{1}{6})$    & $(3,2,\frac{1}{6})$   \\[1ex]
$\chi$    & $( 3,2,-\frac56)$  &  $(1,2,-\frac{1}{2})$ &  $(3,2,\frac{7}{6})$ \\
\bottomrule
\end{tabular}
\hspace{5ex}
\begin{tabular}{cccc}
\toprule
 & $\mathcal{O}_{8}^{4}$ \\
& $(L\bar{d})(\bar{u}^\dagger \bar{e}^\dagger)H$\\
\midrule
$\phi$     & $(3,1,-\frac{1}{3})$\\[1ex]
$\eta$ & $(3,2,\frac{1}{6})$      \\
\bottomrule
\end{tabular}
\caption{Minimal UV completions of operator $\mathcal{O}_8$.}
\label{tab:O8}
\end{table}

\tocless\subsubsection{$\mathcal{O}_{8}^{1}$ Model}
This is the same model as $\mathcal{O}_{3}^{4}$.

\tocless\subsubsection{$\mathcal{O}_{8}^{2}$ Model}
The model contains a complex scalar $\phi$ and a Dirac fermion $\chi+\bar{\chi}$ with quantum numbers
\begin{equation}
\phi\sim\left(3 ,2,\frac16\right),\qquad \chi\sim\left(1,2,-\frac12\right)\;.
\end{equation}
The additional interaction Lagrangian $\Delta\mathcal{L}=\mathcal{L}_Y-\mathcal{V}$ is given by
\begin{align}
-\mathcal{L}_Y&=m_\chi \bar\chi\chi+m_{L\chi,i} L_i \bar\chi+Y^{L\bar d\phi}_{ij} L_i \bar d_j \phi 
+Y^{\bar d\chi\phi}_{i} \bar d_i \chi \phi
+Y^{\bar u\chi\phi}_{i} \bar u_i \bar\chi \phi
+Y^{\bar e\chi H}_{i} \bar e_i \chi \tilde H
+ \hc\\
\mathcal{V}&=\mu^2_{\phi} \phi^\dagger\phi
+ \lambda_\phi(\phi^\dagger\phi)^2
+\lambda_{H\phi} H^\dagger H \phi^\dagger \phi\;.
\end{align}

\tocless\subsubsection{$\mathcal{O}_{8}^{3}$ Model}
The model contains a complex scalar $\phi$ and a Dirac fermion $\chi+\bar{\chi}$ with quantum numbers
\begin{equation}
\phi\sim\left(3 ,2,\frac16\right),\qquad \chi\sim\left(3,2,\frac76\right)\;.
\end{equation}
The additional interaction Lagrangian $\Delta\mathcal{L}=\mathcal{L}_Y-\mathcal{V}$ is given by
\begin{align}
-\mathcal{L}_Y&=m_\chi \bar\chi\chi+Y^{L\bar d\phi}_{ij} L_i \bar d_j \phi 
+Y^{\bar u\chi H}_{i} \bar u_i \chi \tilde H
+Y^{\bar e\bar \chi \phi}_{i} \bar e_i \bar \chi \phi
+ \hc\\
\mathcal{V}&=\mu^2_{\phi} \phi^\dagger\phi
+ \lambda_\phi(\phi^\dagger\phi)^2
+\lambda_{H\phi} H^\dagger H \phi^\dagger \phi\;.
\end{align}

\tocless\subsubsection{$\mathcal{O}_{8}^{4}$ Model}
This is the same model as $\mathcal{O}_{3}^{8}$.

\subsection{Operator $\mathcal{O}_1^\prime=LL\tilde HHHH$}
The only possible non-trivial SU(2)$_L$ structure is
\begin{align}
\mathcal{O}_{1}^\prime=& \left[L L\tilde H\right]_4 \left[HHH\right]_4=L^{\{\alpha} L^\beta\tilde H^{\gamma\}} H^{\{\alpha^\prime}H^{\beta^\prime}H^{\gamma^\prime\}}\epsilon_{\alpha\alpha^\prime}\epsilon_{\beta\beta^\prime}\epsilon_{\gamma\gamma^{\prime}}
\end{align}
where the fields in the brackets $[\dots]_n$ are uniquely contracted to the SU(2)$_L$ representation of dimension $n$. The contraction of three doublets to a quadruplet is symmetric, which is indicated by curly brackets in the last term. The operator has to be completely symmetric under the exchange of the Higgs doublets.

There is one minimal UV completion of the operator $\mathcal{O}_1^\prime=LL\tilde HHHH$ if we do not allow any fields which lead to the usual seesaw models or consider additional Higgs doublets in the external legs. 
The model contains one complex scalar $\phi$ and one Dirac fermion $\chi+\bar\chi$ with 
\begin{equation}
\chi\sim(1,3,-1)\,,\qquad
\phi\sim \left(1,4,\frac32\right)\;.
\end{equation}
This model has been studied in Ref.~\cite{Babu:2009aq}. The additional interaction Lagrangian $\Delta \mathcal{L}=\mathcal{L}_Y-\mathcal{V}$ is given by
\begin{align}
-\mathcal{L}_Y&= Y_{i}^{\chi L H}  \bar\chi \left[L_i \tilde H \right]_3
+ \mu_\chi \bar\chi\chi +Y_{i}^{L\chi\phi} \left[L_i \chi\right]_4 \phi+\hc\\
\mathcal{V}&=\mu_\phi^2 \tilde\phi\phi
+\lambda_{\phi,1} \left[\phi \phi\right]_3 \left[\tilde\phi\tilde\phi\right]_3 
+\lambda_{\phi,2} \left[\phi \phi\right]_7 \left[\tilde\phi\tilde\phi\right]_7 \nonumber \\
&+\left(\kappa \left[H H H\right]_4 \tilde\phi+\hc\right)
+\lambda_{H\phi,1} \left[\tilde H H\right]_1 \left[ \tilde\phi \phi\right]_1 
+\lambda_{H\phi,2} \left[\tilde H H\right]_3 \left[ \tilde\phi \phi\right]_3 
\; . 
\end{align}
There are only two quartic couplings for $\phi$, since the singlet as well as the quintuplet are in the anti-symmetric part of the Kronecker product of two quadruplets.
When $H$ obtains a VEV, the coupling $\kappa$ will automatically lead to a VEV of $\phi$ because of the linear term in $\phi$. 
Thus an experimental constraint from the $\rho$ parameter applies.


\section{Lepton-Flavour Violation}
\label{app:LFV}
We only give the new contributions and refer the reader to Ref.~\cite{Angel:2013hla} for the remaining terms and definitions of the interaction Hamiltonians and matrix elements.

\mathversion{bold}
\subsection{LFV Rare Decay $\mu\to e\gamma$}
\mathversion{normal}

In the limit of a vanishing electron mass, there is an additional contribution to the left-handed part given by 
\begin{equation}
\sigma_{R21}^Y = \frac{m_\mu}{16\pi^2} \frac{Y^{L\bar\chi\phi*}_2 Y^{L\bar\chi\phi}_1}{m_\phi^2} \frac{3+2t-7t^2+2t^3-2(t-4)\ln t}{4(1-t)^4}\;.
\end{equation}

\mathversion{bold}
\subsection{LFV Rare Decay $\mu^- \to e^-e^+e^-$}
\mathversion{normal}
The only new contributions are the additional term in the real photon contribution, the $\gamma$-penguin contribution as well as the box-contribution
\begin{align}
A_1^{Y; L} &= \frac{Y^{L\bar\chi\phi*}_2 Y^{L\bar\chi\phi}_1}{384\pi^2 m_\phi^2} \frac{(t-1)(22+t(13t-41)-2(8-12t+t^3)\ln t)}{(1-t)^4} \;,\\
A_1^{Y;R} &=0 \;,\\
A_2^{Y; L,R}&=\frac{\sigma_{L21,R21}}{m_\mu} \;,\\
B_1^L &= \frac{3 \left(Y_1^{L\bar\chi\phi}\right)^2 Y_1^{L\bar\chi\phi*} Y_2^{L\bar\chi\phi*}}{64\pi^2 e^2 m_\phi^2} \frac{1-t^2+2t \ln t}{(1-t)^3} \;,\\
B_1^R&=B_{2,3,4}^{L,R}=0\;.
\end{align}
There is no new contribution to the $Z$-penguin.
\mathversion{bold}
\subsection{$\mu\leftrightarrow e$ Conversion in Nuclei}
\mathversion{normal}
\label{sec:mu2eConversion}
Similarly to $\mu\to eee$, there is no new $Z$-penguin contribution, but only a contribution to the $\gamma$-penguin and the box diagram. We neglect the box contribution, as it is neglibly small compared to the two penguin contributions:
\begin{align}
g_{LV(d)}^{Y,\gamma} &=-\frac23 \sqrt{2} e^2 A_1^{Y,L} ,&
g_{LV(u)}^{\gamma}=&-2 g_{LV(d)}\;.
\end{align}

\bibliography{draft}

\end{document}